\documentstyle[11pt]{article}

\def\afour{ \setlength{\topmargin}{0mm} \setlength{\headheight}{0mm}
\setlength{\headsep}{0mm} \setlength{\textwidth}{6in}
\setlength{\textheight}{248mm} \setlength{\oddsidemargin}{.25in}
\setlength{\evensidemargin}{.25in} } 

\newcommand\eq[1]{Eq.~(\ref{#1})} \newcommand\eqs[2]{Eqs.~(\ref{#1}) and
(\ref{#2})}

\newcommand\rfrac[2]{\left(\frac{#1}{#2}\right)}

\newcommand{\sub}[1]{_{\mbox{\scriptsize#1}}}
\newcommand{\su}[1]{^{\mbox{\scriptsize#1}}}

\newcommand\ee{\end{equation}} \newcommand\be{\begin{equation}}
\newcommand\eea{\end{eqnarray}} \newcommand\bea{\begin{eqnarray}}

 \newcommand\sunit{\,\mbox{sec}}
 
 \newcommand\km{\,\mbox{km}}

 \newcommand\Mpc{\,\mbox{Mpc}}

\newcommand\mone{^{-1}} \newcommand\mtwo{^{-2}}
\newcommand\mthree{^{-3}} \newcommand\mfour{^{-4}} \newcommand\mhalf{^{-1/2}}
\newcommand\mthreehalf{^{-3/2}} 
 \newcommand\mquarter{^{-1/4}}
\newcommand\half{^{1/2}} \newcommand\threehalf{^{3/2}}
 \newcommand\twothird{^{2/3}}
\newcommand\quarter{^{1/4}}

 \newcommand\mpl{m_{Pl}}

\newcommand\lsim{\mathrel{\rlap{\lower4pt\hbox{\hskip1pt$\sim$}}
    \raise1pt\hbox{$<$}}}
    \newcommand\gsim{\mathrel{\rlap{\lower4pt\hbox{\hskip1pt$\sim$}}
    \raise1pt\hbox{$>$}}}

\newcommand\diff{\mbox d}

 \def\calp{{\cal P}} \def\calr{{\cal R}}

\afour

\newcommand\bfq{\mbox{\bf q}} \newcommand\bfr{\mbox{\bf r}}
 \newcommand\bfw{\mbox{\bf w}}
\newcommand\bfe{\mbox{\bf e}} \newcommand\bfx{\mbox{\bf x}}
 
\newcommand\qperp{\mbox{\bf q}_\perp}

\newcommand\sq{_{\mbox{\scriptsize \bf q}}}
\newcommand\sqp{_{\mbox{\scriptsize \bf q}'}}
\newcommand\msq{_{\mbox{\scriptsize -\bf q}}}

\newcommand\sqperp{\mbox{\scriptsize \bf q}_\perp}

\newcommand\sbfx{\mbox{\scriptsize \bf x}}

  \def\om{\Omega_0}
\def\rt{R^{(3)}}  \def\kl{_{kl}} \def\lm{_{l m}} \def\klm{_{kl
m}}  \def\klm{_{kl m}} 
\def\calr{{\cal R}} \def\rhat{{\widehat \calr}}

\newcommand\cone{  r_1,\theta_1,\phi_1}
\newcommand\ctwo{  r_2,\theta_2,\phi_2}

\begin{document}

\begin{flushright} LANCS-TH/9501\\ astro-ph/9501044\\ \end{flushright}
\begin{center} \Large

{\bf Large scale perturbations in the open universe}

\vspace{.3in} \normalsize \large{David H.  Lyth$^{\dagger}$ and Andrzej
Woszczyna$^*$} \\ \normalsize

\vspace{.6 cm} {\em $^{\dagger}$School of Physics and Materials, \\
University of Lancaster, Lancaster LA1 4YB.~~~U.~K.}\\

\vspace{.2cm}
and

\vspace{.2cm} {\em Isaac Newton Institute, 20 Clarkson Road, Cambridge
CB3 0EH.~~~U.~K.}

\vspace{.4cm} {\em $^*$Astronomical Observatory,
Jagiellonian University,\\
ul. Orla 171, Krakow 30244.~~~Poland. }\\

\vspace{.6 cm} {\bf Abstract} \end{center}

\vspace{.2cm} \noindent

When considering perturbations in an open ($\Omega_0<1$) universe,
cosmologists retain only
sub-curvature modes (defined as eigenfunctions of the Laplacian
whose eigenvalue is less than $-1$ in units
of the curvature scale, in contrast with the super-curvature
modes whose eigenvalue is between $-1$ and $0$).
Mathematicians have known for
almost half a century that all modes must be included
to generate the most general {\em
homogeneous Gaussian random field},
despite the fact that any square integrable {\em function}
can be generated using only the sub-curvature modes.
The former mathematical object, not the latter,
is the relevant one for physical applications.
The mathematics is here explained in a language accessible to
physicists. Then it is pointed out that if the perturbations originate
as a vacuum fluctuation of a scalar field there will be no super-curvature
modes in nature. Finally the effect on the cmb of any super-curvature
contribution is considered, which generalizes to $\Omega_0<1$ the
analysis given by Grishchuk and Zeldovich in 1978.
A formula is given, which is used to estimate the effect. In contrast with
the case $\Omega_0=1$, the effect contributes to all multipoles,
not just to the quadrupole. It is important to find out
whether it has the same $l$ dependence as the data, by
evaluating the formula numerically.

\section{Introduction}

On grounds of simplicity, the
present energy density $\Omega_0$ of the universe is
generally assumed to be equal to unity
(working as usual in units of the critical density).\footnote
{Throughout this article $\Omega_0=1$ will mean a value of $\Omega_0$ close to
1, and   $\Omega_0<1$ will mean a value substantially
less than 1, say less than $0.9$.}
It is not however well determined by
observation \cite{colesellis}. The density of baryonic
matter can only be of order $0.1$ or there will be a conflict with the
nucleosynthesis calculation, and although non-baryonic matter seems to
be required by
observation \cite{dekelrev} there is no guarantee that it will bring the
total up to $\Omega_0=1$.
 Nor should one assume that a cosmological constant
or other exotic contribution to the energy density will play this role.

{}From a theoretical viewpoint the value $\Omega_0=1$ is the most natural,
because any other value of $\Omega$ is time dependent.
 The preference for $\Omega_0=1$ is  sharpened if,  as is widely
 believed, the hot big bang is preceded by an era of inflation.
In that case $\Omega$ has its present value at the epoch when the present
Hubble scale leaves the horizon, and for a generic choice of the inflaton
potential this  indeed implies that $\Omega_0$ is very close to 1
more or less independently of the initial value of $\Omega$.
It is also easier for inflation to explain the homogeneity and isotropy
of the observable universe if $\Omega_0=1$. On the other hand it is
certainly not the case that $\Omega_0=1$ is an unambiguous prediction of
inflation \cite{loomega,lystomega}.

The literature on the  $\Omega_0<1$ cosmology is small compared
with the enormous output on the case $\Omega_0=1$, because the latter is
simpler and observations that can distinguish the two are only now becoming
available.  This is especially true in regard to the subject of the present
paper, which is the effect
of spatial curvature on cosmological perturbations.
  The only data relevant to this subject
are the lowest few multipoles of the cosmic
microwave background (cmb) anisotropy, that were measured recently by the COBE
 satellite \cite{smet,gorski,bunn}.

This article is concerned both with the basic formalism that one should
use in describing cosmological perturbations, and with the cmb
multipoles. To describe its contents, let us
 begin by recalling the presently accepted framework within which
cosmological perturbations are discussed.

  Cosmological perturbations are expanded
in a series of eigenfunctions of the Laplacian for two separate reasons.  One
is that each mode (each term in the series) evolves independently with time,
which makes it easier to evolve a given initial
 perturbation forward in time.  The
other is that by assigning a Gaussian probability distribution
to the amplitude of each mode, one can generate a homogeneous
Gaussian random field. Such a field consists of an ensemble of possible
perturbations, and it is supposed that the perturbation seen in the
observable universe is a typical member of the ensemble.
The stochastic properties of a Gaussian random field are determined
by its two point correlation function $\langle f(1)f(2)
\rangle$, where $f$ is the perturbation and the brackets denote the
ensemble average, and the adjective `homogeneous' indicates that the
correlation function depends only on the distance
between the two points.

The question arises which eigenfunctions to use, and in particular what
range of eigenvalues to include. If $\Omega_0=1$ space is flat
and it is known that the Fourier expansion, which includes all negative
eigenvalues, is the correct choice. It is complete in two distinct
respects. First, it gives the most general {\em square integrable
function}, so that
initial conditions in a finite region of the universe can be evolved forward
in time. Secondly, it gives the most general
{\em homogeneous Gaussian random field}. Instead of the Fourier
expansion one can use the entirely equivalent expansion in
spherical polar coordinates.

 If $\Omega_0<1$,  the curvature of space  defines a
length scale. The spherical coordinate expansion can still be used, and
it is known \cite{fock,bander}
 that the modes which have real negative
eigenvalue {\em less than  $- 1$ in units of the curvature scale}
 provide a complete orthonormal basis for square integrable
functions.
Presumably for this reason, only these modes have been
retained by cosmologists.
We will call them {\em sub-curvature} modes, because they vary
significantly on a scale  which is less than the curvature
scale. The other modes, with eigenvalues between $-1$ and 0 in units of
the curvature scale, we will call super-curvature modes.

It is certainly enough to retain only sub-curvature modes if all one
wishes to do is to
  track the evolution of a given initial perturbation, since
the region of interest is always going to be finite and any function defined
in a finite region can be expanded in terms of the sub-curvature
modes.
(In fact, to describe the observations that we can make it is enough
to specify initial conditions within our past light cone.)
But this is not what one does in cosmology.\footnote
{The only case where one is interested in evolving a given initial condition
 is when one obtains
 the well known relation
between the matter density contrast and the galaxy
peculiar velocity field, but one uses this prediction
 only on very small scales where the curvature cannot be significant.
Even then, one is still interested in the stochastic properties as well.
}
Rather, one uses the mode expansion to
generated a Gaussian perturbation,
by assigning a Gaussian probability distribution to the amplitude of each
mode. In this context  the inclusion of only sub-curvature modes
looks  restrictive.  For example, it leads to  a correlation function
which necessarily becomes small at distances much bigger than the
curvature scale (to be precise, it is less than $r/\sinh r$ times its
value at $r=0$, where $r$ is the distance in curvature units).

Faced with this situation, we queried the assumption that only
sub-curvature modes should be included, and the results of our
investigation are reported here.

First we describe the
mathematical situation, showing that indeed a more general Gaussian
random field is generated by including also the super-curvature modes.
As  expected the correlation function can now be constant out to
arbitrarily large distances.

Then we go on to  ask whether nature has chosen to
use the super-curvature modes, focussing on the low multipoles of the cmb
anisotropy which are the only relevant observational data, and on
the curvature perturbation
which is  thought
to be responsible for these multipoles.
If, as is usually supposed,
this perturbation originates as a vacuum fluctuation of the inflaton
field, there will be no super-curvature modes.
On the other hand, like any other statement about the universe
one expects this assumption to be at best approximately valid.
Supposing that it fails
badly on some very large scale, {\em but that the curvature perturbation still
corresponds to a typical realization of a homogeneous
Gaussian random field}, one is lead
to ask if a failure of the assumption
could be detected by observing the
cmb anisotropy. We note that for $\Omega_0=1$ this question has already
been discussed by  Grishchuk and Zeldovich
\cite{grze},  and we extend their discussion to the case
$\Omega_0<1$.

After our investigation was complete, and the draft of this paper was
almost complete, M. Sasaki suggested to one of us (DHL) that a mathematics
paper written by Yaglom in 1961 \cite{yaglom} might be relevant.
{}From this paper we learned that the need to include both sub- and
super-curvature modes in the expansion of a homogeneous Gaussian
random field in negatively curved space has been known to
mathematicians since at least 1949 \cite{krein}. It would appear
therefore that the assumption by cosmologists that only the sub-curvature
modes are needed is a result of a complete failure of communication
between the worlds of mathematics and science, which has persisted
for many decades. We have retained the mathematics part of our
paper because it gives the relevant results in the sort of language
that is familiar to physicists, though it is strictly speaking
redundant.

Let us  end this introduction by saying a bit more about the
cosmology literature.
Starting with the paper of Lifshitz in 1946 \cite{lifs}, there are
many papers on the treatment of
cosmological perturbations for the case $\Omega<1$.
However, most of them deal with the
{\em definition} and {\em evolution} of the perturbations, which is
 not our main concern.
  We have not attempted
a full survey of this part of the literature, but have just cited useful
papers that we happen to be aware of.
 By contrast, the cosmology literature on
{\em stochastic properties} is very
small for the
 case $\Omega_0<1$, and as we have mentioned it is
out of touch
with the relevant pure mathematics literature where the theory of random
fields is discussed. The first
 serious treatment of stochastic properties
is by Wilson in 1983
\cite{wilson}. He developed the theory from scratch, and not surprisingly
included only the sub-curvature modes which he knew were
sufficient for the description of the non-stochastic properties.
 His notation is defective and much is left
unsaid, but subsequent papers have not made basic advances in the
formulation of the subject, though they have gone much further in
calculating the cmb multipoles and comparing them with observation.
We believe our referencing to be reasonable complete,
as far as the cosmology literature on  the stochastic properties is
 concerned.

The layout of this paper is as follows.  In Section 2
 some basic formulas are given for the Robertson-Walker universe
with $\Omega<1$. In Section 3 the
 standard procedure
is described, and in the next section it is extended to the
super-curvature modes. Inflation is discussed in Section 5, and the
cmb anisotropy is treated in Section 6.
 In an  Appendix we  give various  mathematical results
in the sort of language  that is  familiar to us  as physicists.

\section{Distance scales}

Ignoring perturbations, the universe is homogeneous and
isotropic.  There is a universal scale factor $a(t)$, with $t$ the universal
time measured by the synchronized clocks of comoving observers, and the
distance between any two such observers is proportional to $a$.

According to the Einstein field equation, the time dependence of $a$ is
governed by the Friedmann equation which may be written \be 1-\Omega= -\frac
K{(aH)^2} \label{omega} \ee Here $K$ is a constant, $H=\dot a/a$ is the
Hubble parameter, and $\Omega$ is the energy density measured in units of the
critical density $3H^2/8\pi G$.  (As usual we set $c=1$, and regard any
cosmological constant as a contribution to the energy density as opposed to a
modification of the Einstein field equation.)  The spatial curvature scalar
is \be R^{(3)}=6K/a^2 \label{rthree} \ee The distance $a/|K|$ defines the
curvature scale; on much smaller scales space is practically flat, whereas on
much bigger scales the effect of curvature is very important.
 From \eq{omega} the Hubble
distance $1/H$ is a fraction $(1-\Omega)\half<1$ of the curvature scale.
Even in the extreme case $\Omega_0=0.1$ this makes the curvature scale
 about three times the  present Hubble distance.

  We will set
 $K=-1$ so that $a$ is the curvature scale. Then  the
 case $\Omega=1$ corresponds to the limit $a\to\infty$, with physical
 distances like  $H^{-1}$
 remaining constant.  Note that the effect of curvature
 in a comoving region becomes neither more nor less important with the passage
 of time, since the curvature scale expands with the universe.

 We are concerned with the comoving region which is now the observable
 universe, bounded by the surface of last scattering of radiation emitted at
 very high redshift.  This is close to the particle horizon of a matter
 dominated cosmology, unless there is a cosmological constant or some other
 non-standard contribution to the energy density.  The coordinate distance of
 this particle horizon \cite{kotu} (ie., its distance in units of the
 curvature scale) is $  r\sub{ph}$ where
 $\sinh^2\frac12  r\sub{ph}=\om\mone-1$.  Even the smallest conceivable value
 $\Omega_0\simeq0.1$ gives $  r\sub{ph}=3.6$, so effect of curvature is
 negligible except on scales comparable with the size of the observable
 universe.

 From \eq{omega}, the physical distance of the particle horizon is \be a_0
   r\sub{ph}=(1-\Omega_0)\mhalf H_0\mone   r\sub{ph} \ee For $\Omega_0=1$ it
 is $2H_0\mone$, and even for $\Omega_0=0.1$ it is only $3.8H_0\mone$.  Thus
 it is not very much bigger than the Hubble distance $H_0\mone$.

\section{Sub-curvature modes}

We are concerned with the first order treatment of cosmological
perturbations.  To this order, the perturbations `live' in unperturbed
spacetime, because the distortion of the spacetime geometry is itself a
perturbation.

The perturbations  satisfy linear partial differential equations, in which
 derivatives  with respect to comoving coordinates  occur only through the
Laplacian. When the perturbations
are expanded in eigenfunctions of the Laplacian with eigenvalues
$-(k/a)^2$, each mode (term in the expansion) decouples.

Denoting  the eigenvalue  by $-(k/a)^2$,
it is known  \cite{fock,bander} that the modes with real $k^2>1$ provide
 a complete orthonormal basis for $L^2$  functions, and the  usual procedure
is to keep only
them. Since they all vary appreciably on scales less than the
curvature scale $a$ we will call them sub-curvature modes.
It will be useful to define the quantity
\be
q^2=k^2-1
\ee

\subsection{The spherical expansion}

Spherical coordinates are defined by the line element
 \be \diff l^2 =a^2[ \diff r^2+\sinh^2  r(\diff\theta^2 +\sin^2\theta
\diff\phi^2)]
 \label{elev} \ee
  In the region $  r\ll1$ curvature is negligible and this
 becomes the flat-space line element written in spherical polar coordinates.
The  volume element between adjacent spheres is $4\pi\sinh^2  r\diff   r$,
so for $  r\gg  1$ the  volume $\cal V$ and area $\cal A$
 of a sphere are related by ${\cal V}={\cal A}/2$.
In contrast  with the flat-space case this relation is independent of
$  r$, because most of the volume of a very large sphere is near its
surface.

Since the spherical harmonics $Y_{lm}$ are a complete set on the
sphere, any eigenfunction can be expanded in terms of them. The
radial functions  depend only on $  r$,
and they  satisfy a second order differential equation. As in the
 flat-space case, only one of the two solutions is well behaved at the
origin, so the radial functions are completely determined up
to normalisation. The  mode
expansion of a generic perturbation $f$ is therefore of the form
\be
f(  r,\theta,\phi,t)=\int_0^\infty dq \sum\lm f\klm(t)
Z\klm(  r,\theta,\phi) \label{twen} \ee
where
\be
Z\klm=\Pi_{kl}(  r) Y_{lm}(\theta,\phi)
\ee

 A compact expression for the radial functions is
\cite{dolgov,lifs,vilenkin,bander,harrison}
\be
\Pi\kl=\frac{\Gamma(l+1+iq)}{\Gamma(iq)}\sqrt\frac{1}{\sinh   r}
P^{-l-\frac12}_{iq-\frac12}(\cosh  r)
\label{pileg}
\ee
which  corresponds to the normalisation
\be
\int^\infty_0 \Pi_{kl}(  r) \Pi_{k'l'}(  r) \sinh^2  r \diff  r
=\delta(q-q') \delta_{ll'} \hspace{5em}  \label{xnorm} \ee
The corresponding normalisation of the eigenfunctions is
 \be \int Z^*_{klm} Z_{k'l'm'}
\diff{\cal V}=\delta(q-q') \delta_{ll'} \delta_{mm'} \label{ortho} \ee where
\be \diff{\cal V}=\sinh^2  r\sin\theta d  r d\theta d\varphi
\label{volume} \ee is the volume element.

As it stands \eq{pileg} has a constant nonzero phase. It is convenient
to drop this phase so that the function is real, and one then
has the explicit expressions
\cite{harrison}\footnote {These expressions correct some misprints in
\cite{bida,lystomega}.}  \bea \Pi_{kl} &\equiv& N_{kl} \tilde \Pi_{kl}\\
\tilde \Pi_{kl} &\equiv& q\mtwo (\sinh   r)^l\left(\frac{-1}{\sinh  r}
\frac{d}{\diff  r}\right)^{l+1} \cos(q  r) \label{xdef}\\
N\kl &\equiv& \sqrt\frac2\pi q^2 \left[ \prod_{n=0}^l (n^2+q^2)
\right]\mhalf
\eea
The un-normalised  radial functions $\tilde\Pi_{kl}$
satisfy a recurrence relation \cite{fabbri}
\be
\tilde \Pi_{k ,l+2}=-\left[ (l+1)^2 + q^2 \right]
\tilde \Pi_{k l}+ (2l+3)\coth   r \tilde \Pi_{k ,l+1}
   \label{recur}
\ee
and the first three functions are
\begin{eqnarray}
\tilde \Pi_{k 0}&=& \frac{1 }{\sinh   r} \left[ \frac{\sin (q   r) }{q}
\right] \label{kzero}\\
\tilde \Pi_{k 1}&=&\frac{1 }{\sinh   r} \left [-\cos (q   r)
+\coth  r {\sin (q   r) \over {q}} \right] \\
\tilde \Pi_{k 2}&=&\frac{1}{\sinh   r} \left [
  -3\coth   r \cos (q   r )+(3\coth ^2   r -q ^2 -1)
  {\sin (q   r) \over {q}} \right]
\end{eqnarray}

 The case
$\Omega=1$ corresponds to $q\to \infty$ with $q  r$ fixed, and in that
limit $\Pi_{kl}(  r)$ reduces to the familiar radial function,
\be
\Pi_{kl}(  r)\to \sqrt\frac{2}{\pi}
qj_l(q  r)
\label{jlim}
\,.
\ee
  Near the origin $\Pi_{kl}(  r)$ has
the same behaviour as $j_l(q  r)$, namely $\Pi_{kl}\propto   r ^l$, which
ensures that the Laplacian is well defined there.  The other linearly
independent solution of the radial equation, which corresponds to the
substitution $\cos(q  r)\to\sin(q  r)$ in \eq{xdef}, has the same behaviour
as the other Bessel function $h_l(q  r)$ and is therefore excluded.

\subsection{Stochastic properties}

We are interested in the stochastic properties of the perturbations, at fixed
time.  To define them we will take the approach of considering an ensemble of
universes of which ours is supposed to be one.

The stochastic properties of a generic perturbation $f(  r,\theta,\phi)$ are
defined by the set of probability distribution functions,
 relating to the outcome of a simultaneous measurement of a
perturbation
 at a given set of points.   From the
probability distributions one can calculate ensemble expectation values, such
as the correlation function for a pair of points $\cone$ and $\ctwo$, \be
\xi_f\equiv \langle f(\cone), f(\ctwo) \rangle \label{corr} \ee and the mean
square $\langle f^2(  r,\theta,\phi) \rangle$.

If the probability distributions depend only on the geodesic distances between
the points, the perturbation is said to be {\em homogeneous}
with respect to the group of transformations that preserve this  distance.
(For flat space this is the group of translations and rotations, and
for homogeneous negatively curved space it is isomorphic to the
Lorentz group \cite{gelfand}.)
Then the
correlation function depends only on the distance between the points, and the
mean square is just a number.

Cosmological perturbations are assumed to be homogeneous, and
except for the curvature perturbation that we discuss in Section 6
their correlation functions are supposed to be very small beyond some maximum
distance, called the correlation length.

\subsubsection*{An ergodic universe?}

If there is a finite correlation length,
one ought to be able to dispense with the concept of an ensemble
of universes,
in favour of the concept of sampling our own universe at different locations.
In this approach one defines the probability distribution for
simultaneous measurements at $N$ points
 with by considering  random locations of these points,
 subject to the condition that the distances between them are
fixed. The correlation
function is defined by averaging over all pairs of points a given distance
apart, and the mean square is the spatial average of the square.  For a
Gaussian perturbation in flat space this `ergodic' property can be proved
under weak conditions \cite{adler} and there is no reason to think that
spatial curvature causes any problem though we are not aware of any
literature on the subject.

For the ergodic  viewpoint to be useful, the observable in question has to be
measured in a region that is big compared with the correlation length.
This is the case for  the distributions and peculiar velocities of galaxies
and clusters, where surveys have been done out to several hundred Mpc
to be compared with a correlation length of order $10\Mpc$, and
accordingly the ergodic viewpoint is always adopted there
\cite{peebles}. However, even a distance of a
few hundred  Mpc is only ten percent
or so of the Hubble distance $H_0\mone$, and therefore at most a few percent of
the curvature scale $(1-\Omega_0)\mhalf H_0\mone$. Thus galaxy
and cluster  surveys do not probe spatial curvature.
The only  observables that do, which are the
low multipoles of the cmb anisotropy, are measured only at our position so
there is no practical advantage in going beyond the concept of the ensemble
even if the mathematics turns out to be straightforward.

In addition to
the interpretation that the ensemble corresponds to different
locations within the smooth patch of the universe that we inhabit,
there are two other possibilities.
One is that the ensemble corresponds to different
smooth patches, which are indeed supposed to exist both in
`chaotic' \cite{chaotic} and bubble nucleation
\cite{bubble,neil,misaonew,neilnew}
scenarios of inflation. The other,
adopting the
usual
language of quantum mechanics, is to regard the ensemble
as the set of all possible outcomes of
a  `measurement'
performed on a given state vector.  A concrete realization of this
`quantum cosmology' viewpoint is provided by the hypothesis that
 the perturbations originate as a
vacuum fluctuation of the inflaton field, which we consider later.

\subsection{Gaussian perturbations}

It is generally assumed that cosmological perturbations are Gaussian, in the
regime where they are evolving linearly.  A Gaussian perturbation is normally
defined as one whose probability distribution functions
 are multivariate Gaussians
\cite{karlin,adler,bbks}, and its stochastic properties are completely
determined by its correlation function.  The perturbation is homogeneous
if the correlation function depends only on the distance between the
points.

The simplest Gaussian perturbation is just a coefficient times a given
function, the coefficient having a Gaussian probability distribution.  A more
general Gaussian perturbation is a linear superposition of functions
\cite{karlin},
\be
f(  r,\theta,\phi)=\sum_n f_n X_n(  r,\theta,\phi)
\ee
with each coefficient having an independent Gaussian distribution.
Its stochastic properties are completely determined by the
mean squares $\langle f_n^2 \rangle$  of the coefficients.
(For the moment we are taking the expansion functions $X_n$ to be
real, and to be labelled by a discrete index.)

The correlation function corresponding to the above expansion is
\be
\langle f(\cone) f(\ctwo) \rangle = \sum_n \langle f_n^2 \rangle
X_n(\cone) X_n(\ctwo)
\ee
For it to depend only on the distance between the points requires
very special choices of
the expansion functions, and of the mean squares $\langle f_n^2 \rangle$.

It is very important
to realise that the functions in such an expansion need not be linearly
independent.  Suppose for example that $X_3=X_1+X_2$, and that
$\langle f_3^2 \rangle$ is much bigger than $\langle f_1^2 \rangle$
and $\langle f_2^2 \rangle$. Then most members of the ensemble are
of the form $f=$const$X_3$, which would clearly not have been the case
if the function $X_3$ had been dropped because of its linear
dependence.

So far all our considerations have been at a fixed time.
The time dependence is trivial if we expand in
eigenfunctions of the Laplacian, because each coefficient
$f_n$ then evolves independently of the others.
Let us therefore  replace the
discrete, real expansion above by the complex, partially continuous
expansion \eq{twen}.
The coefficients now satisfy the reality condition $f\klm^*
=f_{kl-m}$, and a Gaussian perturbation is constructed by
assigning independent Gaussian probability distributions to
the real and imaginary parts of the coefficients with
$m\geq 0$. We demonstrate in the Appendix that
the correlation function being dependent only on the distance between
the points is equivalent to the mean
squares of their real and imaginary parts being equal,
and independent of $l$ and
$m$. One can therefore define the
 {\em spectrum} of a generic perturbation
$f$ by
 \cite{lystomega}
 \be \langle f^*_{klm} f_{k'l'm'}\rangle =
 \frac{2\pi^2}{q(q^2+1)} \calp_f(k) \delta(q-q')
\delta_{ll'} \delta_{mm'} \label{29} \ee
The  $q$ dependence of this expression
 has been chosen to give the simple form \eq{msq} for the mean square
perturbation.
In the flat-space limit $q\to\infty$ it reduces to
 \be \langle f^*_{klm} f_{k'l'm'} \rangle =
\frac{2\pi^2}{k^3} \calp_f(k) \delta(k-k') \delta_{ll'} \delta_{mm'}
\label{flatspec} \ee

The
correlation function is given
by
\be \xi_f = \int^\infty_0 \diff q
\frac{2\pi^2}{q(q^2+1)} \calp_f(k) \sum\lm Z^*\klm(\cone)
Z_{klm}(\ctwo)
\label{corr4} \ee
Taking one of the points to be at the origin, only the
 term $l=m=0$ survives,
and using
\be
Z_{k00}(  r,\theta,\phi)=(2\pi^2)\mhalf q\sin(q  r)
/(q\sinh  r)
\ee
one finds
\be
\xi_f(  r)=\int^\infty_1 \frac{\diff k}{k}
\calp_f(k) \frac{\sin(q  r)}{q\sinh  r}
\label{corr2} \ee
Note the appearance in this expression of the logarithmic interval
\be
\frac{\diff k}{k}=\frac{q\diff q}{1+q^2}
\ee
The flat-space limit is
$q\to \infty$ with $q  r$ fixed, and the correlation function then
reduces to
\be \xi_f(  r)=\int_0^\infty \calp_f(k) \frac{\sin(k  r)}{k  r} \frac{\diff
k}{k}
\label{corrflat} \ee
(There is no distinction between $k$  and $q$ in the flat-space
limit, and whenever we consider that limit we will use the
 the  symbol $k$.)
Setting $  r=0$ gives the mean square value
\be
\xi_f(0)\equiv\langle f^2 \rangle =
\int^\infty_1 \frac{\diff k}{k} \calp_f(k)
\label{msq} \ee
The flat-space limit is
\be
\xi_f(0)\equiv\langle f^2 \rangle =
\int^\infty_0 \frac{\diff k}{k}\calp_f(k)
\label{msqflat} \ee

Since the spectrum is positive and $|\sin(q  r)|<q  r$, the
flat-space correlation function is never bigger than its value
at $  r=0$, but  without knowing the spectrum one cannot
say more. The situation is very different for the curved-space
expression \eq{corr2}, because it contains an extra factor
$  r/\sinh  r$. It follows that
 by expanding a perturbation in terms of sub-curvature modes
one obtains a correlation function
bounded by
\be
\frac{\xi_f(  r)}{\xi_f(0)}<\frac{  r}{\sinh   r} \ee

In order  for
$\langle f^2 \rangle$ to be well defined, the spectrum must have appropriate
behaviour at $q=\infty$ and $0$. As $q\to\infty$
 one needs $\calp\to0$.
 As $q\to0$ one needs $\calp\to 0$ in
 the flat case, but only  $q^2\calp_f(k)\to 0$  in the  curved case.

Note that in the curved case the limit $q\to 0$ does {\em not}
correspond to infinite large scales, but rather to scales of order
the curvature scale. This means that one
 cannot tolerate a divergent
behaviour there (unless of course  the curvature  scale happens to be
 larger than any relevant scale,
 in which case we are back to flat space).

For future reference, we note that most other authors have used a
different definition of the spectrum. This is usually denoted by
$P_f$, and it is related to our $\calp_f$ by
\be
\calp_f(k)=\frac{q(q^2+1)}{2\pi^2} P_f(k)
\label{33}
\ee
With this definition,
\be
\xi_f(  r)=\frac{1}{2\pi^2}\int^\infty_0 \diff q q^2
P_f(k) \frac{\sin(q  r)}{q\sinh  r}
\ee

\section{Including the super-curvature modes}

In the last section we found that the usual procedure, which includes
only sub-curvature modes, generates a Gaussian perturbation whose
correlation function necessarily falls off faster than
$  r/\sinh  r$. This reflects the fact that each
super-curvature mode varies strongly  on a scale no bigger than the
curvature
scale. A random superposition of such modes will hardly ever be nearly
  constant on a scale much bigger than the curvature
scale \cite{wilsonfirst}, which is precisely what
the lack of correlation on large scales is telling us.

Faced with this situation one can might think that
the lack of
correlation
on super-curvature scales  is just a mathematical fact,
inherent in the nature of homogeneous negatively
curved space. Certainly it is not trivial to construct a function
exhibiting super-curvature correlations. Consider, for instance,
the following construction; throw down randomly spheres with
radius much bigger than the curvature scale and fill them uniformly
with galaxies, leaving the rest of the universe empty. Then one
might think that a typical observer will see a uniform distribution of
galaxies out to a distance much bigger than the curvature scale,
making the correlation function almost flat out to such a distance.
But this is incorrect, because according to the line element
\eq{elev} most of the volume of a sphere is near its edge, and so is
a typical observer.\footnote
 {One of us (DHL) is indebted to R.  Gott and P.  J.
E.  Peebles for pointing out this fact.}

This example notwithstanding, correlation on arbitrarily large scales
{\em is } possible, and is achieved simply by including the
super-curvature modes.

For $-1<q^2<0$ the analytic continuation of the radial function
$\Pi_{kl}$ is purely imaginary, and for convenience we drop the $i$
factor. Thus the super-curvature modes are defined by
\bea
\Pi_{kl}&\equiv & N_{kl}\tilde\Pi_{kl}\\
\tilde \Pi_{kl} &\equiv& |q|\mtwo (\sinh
  r)^l\left(\frac{-1}{\sinh  r} \frac{\diff}{\diff  r}\right)^{l+1}
\cosh(|q|  r) \\
N_{k0} & \equiv &\sqrt\frac2\pi |q| \\
N\kl &\equiv& \sqrt\frac2\pi |q| \left[ \prod_{n=1}^l
(q^2+n^2) \right]\mhalf \hspace{5em} (l>0)\eea
 The recurrence relation
\eq{recur} is still satisfied, and the first three
functions are
\begin{eqnarray}
\tilde \Pi_{k 0}&=& \frac{1 }{\sinh   r} \left [\frac{\sinh (|q|   r) }
{|q|}\right] \\
\tilde \Pi_{k 1}&=&\frac{1 }{\sinh   r} \left [-\cosh (|q|   r)
+\coth  r {\sinh (|q|   r) \over {|q|}} \right]  \\
\tilde \Pi_{k 2}&=&\frac{1}{\sinh   r} \left [
  -3\coth   r \cosh (|q|   r )+(3\coth ^2   r -q ^2 -1)
  {\sinh (|q|   r) \over {|q|}} \right]
\end{eqnarray}
At large $  r$ the super-curvature modes go like
$\exp[-(1-|q|)  r]$.
Because the volume element is $\diff{\cal V}=\sinh^2  r\sin\theta d  r
d\theta d\varphi $ the integral over all space of a product of any two of
them diverges.  As a result they are not orthogonal in the sense of
\eq{xnorm}, let alone orthonormal.  In any finite region of space (and of
course we are only going to do physics in such a region) they are not even
linearly independent of the sub-curvature eigenfunctions, since the latter
are complete (for the set of $L^2$ functions defined over all space).
None of this matters for the purpose of generating a Gaussian
perturbation.

The super-curvature modes add an additional term to the
expansion \eq{twen},
 \be
f\su{SC}(  r,\theta,\phi)= \int_0^1 d(iq) \sum\lm
f\klm Z\klm(  r,\theta,\phi) \label{twena} \ee
Let us define the corresponding spectrum by
analogy with
\eq{29},
\be \langle f_{klm} f^*_{k'l'm'}
\rangle = \frac{2\pi^2}{|q|(q^2+1)} \calp_f(k) \delta(|q|-|q'|) \delta_{ll'}
\delta_{mm'} \hspace{5mm}(-1<q^2<0)\label{29a} \ee
We show in the Appendix that the correlation function remains well
defined, and dependent only on the distance between the two
points. Taking one of them to
be at
the origin only the $l=0$ mode survives and the
 super-curvature contribution to the correlation
function is  seen to be
\be \xi\su{SC}_f(  r)=
\int_0^1  \frac{\diff k}{k}
\calp_f(k)
\frac{\sinh(|q|  r)} {|q|\sinh  r}
 \label{corr3}
\ee
The super-curvature contribution to the mean square is
\be
\langle f^2 \rangle\su{SC} =
\int_0^1  \frac{\diff k}{k}
\calp_f(k)
\label{mssc}
\ee

\subsection*{Unified expressions including all modes}

The use of $q$ in the mode expansion \eq{twen} is natural for the
sub-curvature modes, and we are using in this paper to facilitate
comparison with existing literature. Unified expressions including
all modes on an equal footing would use $k$ in the mode expansion,
so defining new coefficients $\tilde f\klm$. One would then have the following
expressions, which include both sub- and super-curvature modes.
\bea
f(  r,\theta,\phi,t)&=&\int_0^\infty \diff k \sum\lm \tilde f\klm(t)
Z\klm(  r,\theta,\phi) \\
\langle \tilde f^*_{klm} \tilde f_{k'l'm'}\rangle &=&
 \frac{2\pi^2}{k|q^2|} \calp_f(k) \delta(k-k')
\delta_{ll'} \delta_{mm'}
\label{fullspec}\\
\xi_f(  r)&=&\int^\infty_0 \frac{\diff k}{k}
\calp_f(k) \frac{\sin(q  r)}{q\sinh  r}
\eea

\subsection{Very large super-curvature scales}

The contribution to the correlation function from a mode with
$k^2\equiv 1+q^2\ll1$ is
\be
\xi_f(  r)\propto \exp(-k^2  r)
\label{scasymp}
\ee
for $  r\gg 1$. Thus the correlation length, in units of the curvature
scale $a$, is of order $ k\mtwo$.
This is in contrast with the flat-space case, where
the contribution from a  mode with $k\ll 1$
gives a correlation length of order
$1/k$.
The  difference can be understood in terms
of the different behaviour of the volume element, in the following way.
In both cases, the $  r$ dependence is that of the $l=0$ mode, and
as long as $  r$ is small enough that the mode is approximately constant
the divergence theorem gives
\be
\frac{  r}{f}\frac{\diff f}{\diff    r}
\simeq\frac{-k^2  r  {\cal V}(  r)}{ {\cal A}(  r) }
\ee
where $\cal V$ is the volume within a sphere of radius $  r$
and $\cal A$ is the area of this sphere. In flat space  the
right hand side is equal to  $-(k  r)^2/3$, so
it is  small out to a distance $  r\sim 1/k$. In curved  space,
when $  r\gg  1$, the line element \eq{elev} shows  that
most of the volume of the  sphere is  near its edge, and the right
hand side becomes   equal to $-k^2  r/2$, which is small out to a distance
$  r\sim k\mtwo$.

In addition to being significant in its own right, the correlation
function determines other physically significant quantities.
One, which is relevant for quantities like the density perturbation,
is  the $\sigma^2_f(  r)$, the  mean square of $f$ after it  has been
smeared over a sphere of radius $  r$,
 which is given by
\be
 \sigma_f^2(  r)={\cal V}(  r)\mtwo\int\diff{\cal V}_1
\int \diff {\cal V}_2
\xi_f(\cone,\ctwo)
\ee
where the integrations are within the spheres $  r_1<  r$
and $  r_2<  r$ (we have taken advantage of
homogeneity to  evaluate locate the sphere at the origin).
Because most of the volume of the sphere is near it's edge,
the behaviour \eq{scasymp} of the correlation  function  leads to
the same behaviour for  $\sigma_f^2(  r)$, so it too remains constant
out to a distance $  r\sim k\mtwo$. Another  quantity, which is relevant
for  the curvature  perturbation $\calr$
 that we shall discuss
in  Section 6, is  the mean square after smearing over
 a geodesic surface of  radius $  r$,
\be
{\cal A}(  r)\mtwo
\int\diff{\cal A}_1\int\diff {\cal A}_2 \xi_\calr(\cone,\ctwo)
\ee
On  super-curvature scales it is a measure of the fractional
perturbation in the curvature of the surface. One sees  that
 it too  shares the behaviour
\eq{scasymp}, and so is  constant out  to a distance $  r
\sim k\mtwo$.

\section{The cmb anisotropy}

Even if $\Omega_0<1$, spatial curvature is negligible on scales that are
small compared with the Hubble distance.  As a result, the only
observational  data that can be sensitive  to curvature, even if
$\Omega_0<1$, are
 the lowest few multipoles of the cmb anisotropy.
Papers discussing the effect of curvature on these multipoles
\cite{wilsonfirst,wilson,early,lystomega,suggou,gouda}
appeared sporadically before
they were measured by the COBE satellite
\cite{smet}, and many have appeared since
\cite{kamsper,both,kamspersug,kametal,ratrapeeb,tegsilkopen,kashlinsky}.
All of these papers  keep
only sub-curvature  modes. Here  we consider both sub- and
super-curvature modes.

The multipoles are defined by \be \frac{\Delta
T(\bfe)}{T}=\bfw.\bfe+\sum_{l=2}^{\infty} \sum_{m=-l}^{+l} a_{lm} Y_{lm}
(\bfe) \label{mult} \;.\ee The dipole term $\bfw.\bfe$ is well measured, and
is the Doppler shift caused by our velocity $\bfw$ relative to the rest frame
of the cmb.  Unless otherwise stated, $\Delta T$ will denote only the
intrinsic, non-dipole contribution from now on.

If the perturbations in the universe are Gaussian, the real and imaginary
part of each multipole will have an independent Gaussian probability
distribution (subject to the condition $a^*_{lm}=a_{l,-m}$).  The expectation
values of the squares of the real and imaginary parts are equal so one need
only consider their sum, \be C_l \equiv \left\langle |a_{lm}|^2 \right\rangle
\;.\ee Rotational invariance is equivalent to the independence of this
expression on $m$.

Even if it can be identified with an average over observer positions,
the expectation value $C_l$  cannot be measured.
Given a theoretical prediction for $C_l$, the
best guess for $|a_{lm}|^2$ measured at our position
 is that it is equal to $C_l$, but one can also
calculate the variance of this guess, which is called the {\it cosmic
variance}.  Since the real and imaginary part of each multipole has an
independent Gaussian distribution the cosmic variance of $\sum_m|a_{lm}|^2$
is only $2/(2l+1)$ times its expected value, and by taking the average over
several $l$'s one can reduce the cosmic variance even further.  Nevertheless,
for the low multipoles that are sensitive to curvature it represents a
serious limitation on our ability to distinguish between
different hypotheses about the $C_l$. Any hypothesis can be made consistent
with observation by supposing that the region around us is sufficiently
atypical.

The surface of last scattering of the cmb is practically at the particle
horizon, whose coordinate distance is $\eta_0$ with
$\sinh^2\eta_0/2=\om\mone-1$.  An angle $\theta$ subtends at this surface a
coordinate distance $d$ given by \cite{peebles} \be \theta=\frac12
(1-\om)\mhalf\om d
=\frac12(a_0H_0\Omega_0 d)
\label{subtend}
\ee Spatial curvature is negligible when $d\ll 1$,
corresponding to \be \theta\ll 30(1-\om)\mhalf\om \mbox{\ degrees}
\label{fofo} \ee
A structure with angular size $\theta$\,radians is
dominated by multipoles with \be l\sim1/\theta \label{lofthe} \ee
one expects that spatial
curvature will be  negligible for  the multipoles
 \be l\gg \frac{2\sqrt{1-\Omega_0}}{\Omega_0}
\label{lscale}
\ee
This is the regime
 $l\gg 20$ if $\Omega_0=0.1$, and the regime $l\gg 6$ if
$\Omega_0=0.3$.

This restriction need not apply to super-curvature
modes with $k^2\ll 1$ because the spatial gradient involved is then
small in units of the curvature scale. The contribution of these modes
is called the Grishchuk-Zeldovich effect, and we discuss it later.

The linear scale probed by the multipoles decreases as $l$ increases,
and for $l\sim 1000$ it becomes of order $100\Mpc$. On these scales
one can observe the distribution and motion of galaxies and clusters
in the region around us.
On the supposition that they all have a common origin, the
cmb anisotropy and the motion and distribution of galaxies and clusters
are collectively termed `large scale structure'.

A promising model of large scale structure is that it originates as an
adiabatic density perturbation, or equivalently
\cite{bardeen,ly85,lymu,ellis} as a perturbation in
the curvature of the hypersurfaces  orthogonal to the  comoving
worldlines. This model has has been widely investigated for the case
$\Omega_0=1$ \cite{LL2}, and recently it has been advocated also for
the case $\Omega_0<1$ \cite{kamsper,ratrapeeb,kamspersug}.
In this paper we consider the model only in relation to the cmb
anisotropy since the galaxy and cluster data are insensitive
to spatial curvature. We note though that the full data set may impose
a significant lower bound on $\Omega_0$ \cite{ll94}.

\subsection{The curvature perturbation}

The curvature perturbation is conveniently characterised by a quantity
$\calr$, which is defined in terms of the perturbation in the curvature
scalar by\footnote
{The
quantity $\calr$ was called $\phi_m$ by Bardeen who first considered it
\cite{bardeen},  $\calr_m$ by Kodama and Sasaki \cite{ks84}.
It is equal to $3/2$ times the quantity $\delta K/k^2$ of
Lyth \cite{ly85,lymu}, which is in turn equal to the
$\zeta$ of Mukhanov, Feldman and Brandenberger \cite{mukrev}.
After matter domination it is equal to $-(3/5)\Phi$, where $\Phi$ is the
peculiar gravitational potential (and one of the `gauge invariant'
variables introduced in \cite{bardeen}).
  On scales far outside the horizon, in the case  $\Omega=1$,
 it is the $\zeta$ of
\cite{sb89}, and three times the $\zeta$ of \cite{bs83}. }
 \be 4(k^2+3) \calr\klm/a^2 =\delta \rt\klm \label{twei} \ee
In the limit $\Omega\to 1$,
\be
4k^2\calr\klm/a^2=\delta \rt\klm \label{tweiflat}
\ee

On cosmologically interesting scales,
 $\calr\klm$ is expected to be  practically constant in the
early universe.
To be precise, it is  practically constant on scales far outside the horizon
in the regime where $\Omega(t)$ is close to
1 (assuming that the density perturbation is adiabatic)
\cite{ly85,lymu,ellis,new}. During matter domination
the former condition can be dropped, so that $\calr\klm$ is constant on
all scales until $\Omega$ breaks away from 1. After that it has the
time dependence
$\calr\klm=F\hat\calr\klm$ where $\hat\calr\klm$ is the
early time constant value and
\be
F=5\frac{\sinh^2\eta-3\eta\sinh\eta+4\cosh\eta-4}{(\cosh\eta-1)^3}
\label{rvar}
\ee
with
\be
\eta=2(aH)\mone=2(1-\Omega)\half
\ee
 Unless otherwise indicated, $\calr$ will indicate the primordial
value $\rhat$ from now on.

During matter domination and before $\Omega$ breaks away from
1, the density contrast is given by
\be
\rfrac{a^2H^2}{k^2+3}\frac{\delta\rho\klm}{\rho}=\frac
25 \calr\klm
\label{density}
\ee
For $\Omega_0=1$ this reduces to
\be
\rfrac{a^2H^2}{k^2}\frac{\delta\rho\klm}{\rho}=\frac
25 \calr\klm
\label{densflat}
\ee
In these expressions the density perturbation is evaluated on comoving
hypersurfaces (it is often referred to as the `gauge invariant' density
perturbation). In the matter dominated era where they hold this is
the same as the density perturbation in the `synchronous gauge' with the
`gauge mode' dropped \cite{lyst90}.

\subsection{The Sachs-Wolfe effect}

Horizon entry occurs long after matter domination on scales \be
a_0/k\gg  20(\Omega_0 h^2)\mone \Mpc \ee where $h$ is the
value of $H_0$ in units of $100\km\sunit\mone\Mpc\mone$.
As a  fraction of the Hubble distance these are the scales
 \be
\frac{a_0 H_0}{k}\gg \frac{.007}{\Omega_0 h} \label{vlarge} \ee
It  follows  from \eq{subtend} that they correspond to multipoles
\be
l\ll 300h  \ee
Assuming an initial adiabatic perturbation,these multipoles
 are dominated  by the effect of the
distortion of the spacetime metric between us and the surface of last
scattering, which is called Sachs-Wolfe effect. If
 $\Omega_0=1$ the Sachs-Wolfe approximation accounts for about
90\% of $C_l$
at  $l=10$, and about  $50\%$ at $l=30$ \cite{bunn}.

The Sachs-Wolfe effect is determined by the curvature perturbation.
In the case $\Omega_0=1$ it is given by
 \cite{peebles,LL2,brly}
\be
\Delta T(\bfe)/T=-\frac15\calr(\eta_0\bfe)
\ee
 where $\eta_0=2(a_0H_0)\mone$ is
the coordinate distance of the edge of the observable universe,
taken to correspond to the surface of last scattering.
(The right hand side is usual given as $\frac13\Phi(\eta_0\bfe)$ where
$\Phi$ is the peculiar
gravitational potential.)  Using \eq{jlim}  the
multipoles are therefore given by
 \be
a_{lm}=-\frac15\int^\infty_0\diff k\calr\klm
\sqrt\frac2\pi k j_l(\eta_0 k)
\label{424} \;.\ee
Using \eq{flatspec}, the
mean square multipoles $C_l$ are therefore given by
\be
C_l=
 \frac{4\pi}{25} \int_0^{\infty} \frac{\diff k }{k} \, j_l^2 (\eta_0
k)
\; \calp_\calr(k) \label{clflat} \ee

For $\Omega_0<1$, keeping for the moment only sub-curvature modes,
the Sachs-Wolfe
effect is given by \cite{lystomega,suggou,gouda}
\bea a\lm &=&-\int^\infty_0 dq
 \calr\klm
q I\kl
\label{73}\\
q I\kl&=&\frac15 \Pi\kl(\eta_0) +\frac65 \int^{\eta_0}_0 \diff  r
\Pi\kl(  r)F'(\eta_0-  r)
\label{74}
 \eea
Here $\calr\klm$ is evaluated well before $\Omega$ breaks away from
1 (when it is constant),
$F$ is given by \eq{rvar} and $\eta_0=2(1-\Omega_0)\half$
is again the coordinate distance of the edge of the observable universe.
 The mean square multipoles are given by
\be
C_l=2\pi^2 \int^\infty_1 \frac{\diff k}{k}
\calp_\calr(k) I\kl^2 \label{opensw} \ee
When $k\to 1$,   $I_{kl}$ tends to a finite and nonzero limit for
each $l$. This means that  the
$C_l$ are finite provided that
$q^2\calp_\calr\to 0$.

Two things should be noted about the  regime $q\to 0$ in the curved space
case. First,
all multipoles receive contributions from this regime; in contrast with
the flat case  the quadrupole does not  dominate  as is claimed in
\cite{kashlinsky}). Second,
 the limit $q\to0$  corresponds
to scales of order the curvature, not to infinitely large scales
as is claimed  in \cite{kamsper,kashlinsky}.
Because of this last  fact, one
cannot  tolerate a divergence of the $C_l$ as $q\to 0$
(unless the curvature
scale   is much bigger  than any scale of interest in which case
one is back to   the flat-space case).

\subsection{Inflation and horizon exit}

It is widely supposed that the hot big bang is preceded by an era of
inflation, during which gravity is by definition repulsive.
A very attractive hypothesis is that the curvature perturbation
originates as a vacuum
fluctuation during inflation, so that the ensemble average appearing in the
definition of the spectrum (\eq{29})
is just the vacuum expectation value. Made originally for the case
$\Omega_0=1$   \cite{adpred,bs83,ly85,mukhanov,sasaki},
this hypothesis was later extended to the case $\Omega_0<1$
by Lyth and Stewart \cite{lystomega}.
Before discussing it, let us see how inflation works with special
reference to the case $\Omega_0<1$.

The Hubble distance  $H\mone$ is usually termed the
horizon (to be distinguished from the particle horizon), and
the comoving length scale $a/k$ associated with a given mode
is said to be outside the horizon if $aH/k>1$, and inside
the horizon if $aH/k<1$. The evolution of perturbations outside the
horizon is very simple, because it is not affected by causal processes.
Instead, the perturbation evolves independently in each comoving
region \cite{LL2}.

Super-curvature scales, $a/k>1$, are  always outside the horizon
(from \eq{omega}), but sub-curvature scales can be either outside or
inside it.
In the usual cosmology where gravity
is attractive, $aH\equiv\dot a$  decreases with  time and
at each epoch some scale is entering the horizon.
The  Hubble  scale $H_0\mone$
is entering the horizon now, and smaller scales entered
the horizon earlier.  Also, from \eq{omega}, $\Omega$ is driven away from 1
as time passes, so that $|1-\Omega|$
must have been extraordinarily small at early times even
if it is  not  small now.

Inflation may be defined as an early era of repulsive gravity,
when $aH\equiv\dot a$ increases with time, and it is widely supposed
that such an era preceded the hot big bang. At each epoch
during inflation some scale is leaving the horizon, and as time goes by
$\Omega$ is driven  towards 1.
The standard assumption is that
inflation occurs because the scalar field potential dominates the
energy density, which falls slowly with time owing to the evolution
of one of the scalar fields, termed the inflaton field.
Constant energy density corresponds to
$\Omega\propto H\mtwo$, and combining this dependence
with \eq{omega} gives, for the case $\Omega<1$,
\be
a= \hat H \sinh(  \hat  Ht)
\label{aeq}
\ee
and
\be
H=\hat H \coth(\hat H t)
\label{heq}
\ee
After $\Omega$ has been driven to 1, $H$ achieves the
almost constant value $\hat H$, and
\be
a\propto\exp(H t)
\label{aeqflat}
\ee
It is related to the scalar field potential $V$ (in turn practically
equal to the energy density) by
\be
\hat H^2=\frac{8\pi}{3}\mpl\mtwo V
\label{hath}
\ee
(After many Hubble times the Hubble constant could vary appreciably,
in which case $\hat H$ denotes the value before this happens.)

This evolution of the scale factor is modified if the energy density
is rapidly  decreasing with time; in particular, inflation might
begin with a `coasting epoch' during which
$\Omega$ is almost constant \cite{lystomega,edet}. We shall not
consider that case.

In the case $\Omega_0=1$, inflation is usually held to solve at least
three problems that arise if the hot big bang extends back to the
Planck scale.
Let us briefly recall them, and consider whether they
are still solved if $\Omega_0<1$. For the moment we are discounting the
bubble nucleation model of inflation.

\begin{enumerate}
\item {\em The harmful relic problem}
Without inflation it is difficult to avoid harmful relics of the early
universe like monopoles or gravitinos. The possibility of avoiding them
by inflation does not depend on the value of $\Omega_0$.
\item {\em The flatness problem}
 As we go back
through time during the hot big bang,
$\Omega$ is driven very close to 1 so that we have a fine tuning
problem, the flatness problem. It is solved provided that inflation
lasts long enough that $\Omega$ has been driven away from 1 again
as we go back to the beginning of inflation. From
\eq{aeq} (or its $\Omega>1$ counterpart), the era at the beginning of
inflation during which $\Omega$ is significantly different from 1
has a typical duration of order
$\hat H\mone$ where $\hat H$ is related to the scalar
field potential by \eq{hath}. From \eq{omega}, $\Omega$ has its
present value $\Omega_0$ at the epoch during inflation
when the observable universe (or to be precise, the
comoving scale presently equal to the Hubble distance) leaves the
horizon. Thus, $\Omega_0$ will be different from 1 if this epoch occurs
near the beginning of inflation on the timescale $\hat H\mone$, but
close to 1 otherwise.
\item
{\em The homogeneity (horizon) problem}
Without inflation, the observable universe is far outside the horizon
(Hubble distance) at early times. This
means that causal processes cannot determine
the initial conditions, which is usually held to be a problem, termed
the `horizon problem'.
If $\Omega_0$ is close to 1,
 the observable universe is typically far
inside the horizon at the beginning of inflation, which solves the
horizon problem and is usually said to `explain' the homogeneity of the
observable universe. Inflation with $\Omega_0<1$ cannot solve the
horizon problem because the observable universe
(or to be precise the comoving length presently equal to the Hubble
distance) never occupies less than a fraction $1-\Omega_0$ of the Hubble
distance.

However, no causal mechanism has ever been
proposed for actually establishing homogeneity at the beginning of
inflation, even after the horizon problem has been solved.
It seems to us therefore that the `horizon problem' is a red herring,
and that one should therefore look elsewhere for an explanation of the
homogeneity of the universe. For the case $\Omega_0\simeq 1$
a fruitful avenue seems to be the following \cite{LL2}.
As smaller and smaller scales are considered one
expects to find homogeneity below some minimum scale, but this is not
the Hubble distance even though that is the only scale available at the
classical level. Rather it is the
scale, available only at the quantum level,
\be
\rho\mquarter=\rfrac{3}{8\pi}\quarter
\rfrac{H}{\mpl}\half H\mone
\ee
(we are setting $\hbar$ as well as $c$ equal to 1).
Indeed, within
a volume with this radius, even the vacuum fluctuation of a massless
scalar field generates energy density {\em and pressure} of order
$\rho\quarter$, which would spoil inflation.
 As in the case of flat
spacetime this vacuum contribution to the energy density is to be
discounted (ie., one has to solve the cosmological constant problem
by fiat at our present level of understanding).
But one cannot allow a significant occupation number for the
particle states defined on this vacuum.
In other words, if $\Omega_0=1$
the universe has to be {\em absolutely homogeneous} at the classical level,
on scales smaller than $\rho\mquarter$. This guarantees
the homogeneity of the observable universe
at the classical level, provided that
inflation starts at least $[\ln(\mpl/H_1)]$ Hubble times before
the observable universe leaves the horizon, where $H_1$ is the
value of $H$ at this latter epoch. In order to
respect the isotropy of the cmb one requires
$(H_1/\mpl)\half\lsim 10\mthree$ \cite{lyth84},
and the bound is saturated in typical models of inflation.
Thus, homogeneity of the observable universe is typically guaranteed
if it leaves the horizon more than 7 or so Hubble times after
the beginning of exponential inflation \cite{LL2}.

If $\Omega_0<1$ it is unclear how to define the vacuum as we discuss
below, but with the mathematically simple conformal vacuum the vacuum
fluctuation again generates an energy density and pressure of order
$d\mfour$ on the scale $d$. The criterion that this should not spoil the
inflationary behaviour \eq{aeq} is that $d\mfour$ be much less than the
critical density, which requires as before
$d\gsim (H/\mpl)\half H\mone$. But now the observable
universe is never far inside the Hubble distance $H\mone$, so its
homogeneity is not guaranteed by this type of argument.
A different avenue would be to invoke quantum cosmology, along the lines
of \cite{halliwell} which however deals only with the case
$\Omega_0>1$.
\end{enumerate}

\subsubsection*{The bubble nucleation model of inflation}

All of the above discussion assumes a classical evolution for
the inflaton field, leading to a smooth evolution of $\Omega(t)$.
It might happen, however, that the scalar field potential allows
quantum tunneling in scalar field space at some point during inflation.
In that case a bubble of scalar field can form, whose interior is
an $\Omega\ll 1$ universe \cite{bubble,neil,misaonew,neilnew}.
 Provided that the
scalar field potential is still flat enough, $\Omega$ will again
be driven to 1.

If $\Omega_0$ turns out to be less than 1 the bubble nucleation
model will be very attractive. Homogeneity is automatic. Also,
$\Omega_0$ is determined by the form of the scalar field potential
and can easily be less than 1 \cite{neil}.
Assuming the usual `chaotic' scenario for the beginning of inflation
\cite{chaotic,eternal,linde},
the inflaton field rolls slowly down a valley in scalar field space, and
then the bubble nucleation model might correspond to
sideways tunneling out of this valley \cite{ewan}.
The only problem would be to find a potential of the required
form that looks sensible in the context of modern particle theory;
as has recently been pointed out \cite{edet,ewanstuff},
 this constraint makes it difficult
even to find a potential that leads to ordinary, non-tunneling
inflation.

\subsection{Sub-curvature contributions and the vacuum fluctuation}

During inflation, the curvature perturbation is related to the perturbation
$\delta\phi$ of the inflaton field
 by   \cite{sasaki,lystomega,LL2}
\be
\calr=-(H/\dot\phi)\delta\phi
\label{75}
\ee
where the dot denotes differentiation with respect to time $t$.
This expression  holds at all epochs, not just when $\calr$ is constant.
In it,  $\delta\phi$ is
defined \cite{sasaki} on hypersurfaces which have
zero perturbation in their curvature scalar (it is often called the
`gauge invariant' inflaton field  perturbation).

A very attractive hypothesis is that $\delta\phi$ originates as a vacuum
fluctuation, so that the ensemble average appearing in the
definition of the spectrum (\eq{29}) is just the vacuum expectation value
  \cite{adpred,bs83,ly85,mukhanov,sasaki,stewly}.

The vacuum fluctuation during  inflation also
 generates  a spectrum of gravitational waves,  which
is well understood for the case $\Omega_0=1$ \cite{starobinsky},
and is under investigation  for the case $\Omega_0<1$
 \cite{bruce}. We will not consider it here.

To
calculate the vacuum fluctuation  one uses
quantum field theory in negatively curved space
\cite{bida,wald}, and the first step in
setting
up this theory is to expand  $\delta
\phi$ in terms of the {\em sub-curvature} mode
functions. In this context there is no question of including
additional modes, because many results of quantum field theory
(such as the vanishing of field  commutators outside the light
cone) depend essentially on the fact that one is using a complete
orthonormal
set. As a result the spectrum predicted by the vacuum fluctuation
will include only sub-curvature modes.

The same restriction holds for the fluctuation in any quantum state
that is homogeneous (with respect to the group of coordinate
transformations leaving the distance between each pair of points
invariant). But one can give the inflaton field perturbation
any desired stochastic properties by choosing a suitable
quantum state (pure or mixed), and in particular one can generate
an arbitrary homogeneous Gaussian perturbation.
The absence of super-curvature modes in the vacuum fluctuation
prediction is not a feature of quantum field theory {\em per se}.

The coefficients in the mode expansion of the quantum field
$\phi\klm$ satisfy the classical field equation (in the Heisenberg
representation), which fixes them up to a one-parameter
ambiguity once a convention is made for their normalisation.
Breaking this ambiguity is equivalent to defining the
vacuum.
In the case $\Omega_0=1$ each mode starts out well inside the horizon,
 where the spacetime curvature is negligible. In that case the
vacuum is defined to be the usual flat spacetime vacuum, and assuming the
usual slow roll conditions one finds \cite{ly85}
\be
\calp_\calr(k)\half=\frac{8}{\mpl^3}\sqrt\frac{2\pi}{3}\frac{V\threehalf}
{V'}
\label{ppred}\ee
In this expression $V(\phi)$ is the inflaton potential, and
the right hand side is to be evaluated at the epoch of horizon exit $k=aH$.
It gives an almost scale-independent result for typical models
of inflation.

In the case $\Omega_0<1$, without bubble nucleation, it is not clear how
to define the vacuum because a given scale is never far inside the
horizon. The mathematically simplest choice is the `conformal vacuum',
and using it one finds (after suitably generalising the slow roll
conditions) that $\calp_\calr$ is still given by the above
expression \cite{lystomega,new}. (For the special case of a linear
potential this result has been reproduced recently, using a different
calculational technique \cite{ratrapeeb}.\footnote
{The authors of \cite{ratrapeeb} do
not establish the identity of their result with
the earlier one, but it follows by evaluating Eq.~(2)
of \cite{kamspersug} (multiplied by
$16\pi/\mpl^2$ to bring the conventions of \cite{ratrapeeb} into line
with the usual ones) during matter domination before $
\Omega$ breaks away from unity. To do this one has to
replace $1-\Omega_0$ by $1-\Omega\ll1$ in the quantity
$W_1/c_1$, leading to \cite{peebles} $W_1/c_1\to (2/5)(aH)
\mtwo$. Remembering that
the energy density scales like $a\mthree$ during matter domination
and like $a\mfour$ during radiation domination,  one then indeed reproduces
the spectrum of the energy density given by \eqs{density}{ppred}.
})

In the bubble nucleation model  the quantum state of the inflaton
field perturbation inside the bubble can be calculated
\cite{misaonew,neil,bruce}, and it is found
{\em not} to be in the conformal vacuum.
As a result \cite{neil,neilnew},
$\calp_\calr(k)$ is multiplied by a factor $\coth(\pi q)$ compared with
\eq{ppred}.

\subsubsection*{Comparison of the vacuum fluctuation with observation}

 If $\Omega_0=1$, \eq{clflat}
with a  scale independent  $\calp_\calr$
gives
\be
l(l+1)C_l=\frac{2\pi}{25}\calp_\calr
\ee
The prediction is that the left hand side is scale independent,
which is consistent with observation   \cite{gorski,bunn}.
There is however room for considerable scale dependence;
defining $n$ by $\calp_\calr\propto q^{n-1}$,
the allowed range is $0.6\lsim n\lsim 1.4$.
The magnitude is small,
\be
l(l+1)C_l=8.05\times 10^{-10}\frac{Q\sub{RMS-ps}}{20\,\mu{\mbox K}}
\simeq 8.0\times 10^{-10}
\label{cmag}
\ee
which  corresponds to
\be
\calp_\calr\simeq 3\times 10^{-9}
\label{pmag}
\ee

For $\Omega_0<1$ \eq{opensw} has to be evaluated numerically.
With the flat spectrum coming from the conformal vacuum assumption,
$l(l+1)C_l$ has in general a negative slope for
$0.1<\Omega_0<1$ (for $l$ in the range $l\lsim 10$ where the Sachs-Wolfe
effect dominates, and in which curvature can be significant)
\cite{lystomega,ratrapeeb,kamspersug,gorskiopen}.
With present data the slope is not strong enough to rule out any
value of $\Omega_0$, though better data will probably rule out
low values.

As already noted, the spectrum is not flat in the bubble nucleation
model, but rather is proportional to $\coth(\pi q)$.
It turns out however \cite{neilnew} that if $\Omega_0$ is substantially
below 1 the integral in \eq{74} dominates, with $I\kl^2$
peaking at $q^2\gsim 1$ even for the quadrupole
(and at higher values for higher multipoles). As a result the
bubble nucleation prediction is not significantly different from
the flat spectrum prediction, when cosmic variance is taken into
account.

\subsubsection*{Power law parameterization of the density perturbation
spectrum}

The spectrum $\calp_\calr$ of the curvature perturbation is
directly related to the Sachs-Wolfe effect, and from this viewpoint
the assumption that it is effectively flat seems natural.
This assumption was not, however, the assumption made in the literature
before the (very recent) advent of the vacuum fluctuation prediction.
Rather, it was assumed that the spectrum $P_\delta$ (defined by
\eq{33}) of the density
perturbation is proportional to $q$.
This choice is equivalent to the flatness of $\calp_\calr$ for
$\Omega_0=1$, but otherwise it is equivalent to
\be
\calp_\calr\propto \frac{q^2(1+q^2)}{(4+q^2)^2}
\ee
The right hand side tends to 1 in the limit $q^2\to \infty $ of
negligible curvature, but is much less than 1 for $q^2\lsim 4$.
With this parameterization, $l(l+1)C_l$ acquires a
{\em positive} slope \cite{both} for $0.1<\Omega_0<1$, comparable in
magnitude with the negative slope of the vacuum fluctuation prediction.

The main cause of this difference is the
factor $(4+q^2)^2$ coming from the relation between
the density and curvature perturbations, and a similar result would
probably be obtained if $\calp_\delta$ were used instead of
$P_\delta$, or $k$ instead of $q$. In other words, the predicted
$C_l$ can be regarded as coming simply from a {\em linearly rising
density perturbation spectrum}, as opposed to
a {\em flat curvature perturbation spectrum}. These two
parameterizations
are equivalent if $\Omega_0$ is equal to 1, but if $\Omega_0$
is significantly smaller than 1
the first one gives less power on small scales, leading to
a significantly different prediction for the
$C_l$'s. The central values of the present data points lie between the
two predictions, and the present error bars are big enough that
they are indistinguishable.
But in the future the data should be able to distinguish between the two
parameterizations, ruling out one or both of them for small values
of $\Omega_0$.

\subsection{Super-curvature scales and the Grishchuk-Zeldovich effect}

Like any statement in physics, the statement that the inflaton field
is in the vacuum will be at best approximate, and its validity will
presumably depend on the scale under consideration. When considering
departures from it there is no reason to exclude super-curvature scales.

The contribution of super-curvature scales to the mean square
multipole $C_l$ is  just the  extension of  \eq{opensw}
to  super-curvature scales,
\be
C_l\su{SC}=
2\pi^2 \int^1_0 \frac{\diff k}{k}
\calp_\calr(k) I\kl^2 \label{openswsc} \ee

Requiring that the super-curvature contribution be no bigger than the
total gives (\eq{cmag})
\be
l^2 C_l\su{SC}\lsim 10^{-10}
\label{cvlbound}
\ee

The quantities $I\kl^2$ have not been calculated in
the super-curvature regime $0<k^2<1$, but
as discussed below they
are proportional to $k^2$ near $k=0$, and
we noted earlier that in the regime $k^2>1$ they
all peak at a value $k^2\gsim 2$. It therefore seems reasonably to suppose
that for a fairly flat spectrum
 $\calp_\calr(k)$, the super-curvature
regime does not contribute much to the mean square multipoles $C_l$.
In that case \eq{cvlbound} will provide no significant constraint
on a flat spectrum. In other words, it will be difficult to detect the
cutoff below $k^2=1$ that the vacuum fluctuation predicts.
(The even more difficult task of finding an observational
signal for the this cutoff {\em without} assuming that the spectrum
is flat is discussed in \cite{rob}.)

Now suppose, in contrast, that the spectrum rises sharply on some very
large scale. Then there might be a big curvature perturbation, which
would however have a big correlation length and so have a very small
spatial gradient. If the correlation length is big enough,
the gradient will be small enough to
ensure that the curvature perturbation has no significant effect
on the cmb anisotropy, even if it is quite is large. How big does the
correlation length have to be for this to happen?

For $\Omega_0=1$ this question was
asked and essentially
 answered by Grishchuk and Zeldovich \cite{grze}.
We will now
briefly recall their argument, using the precise  concept of the
  spectrum of the curvature perturbation in place of their more qualitative
discussion. Then we will generalize it to the case $\Omega_0<1$,
and finally discuss its physical significance in both cases.

\subsection{The Grishchuk-Zeldovich effect ($\Omega_0=1$)}

{}From \eq{msqflat}, a flat spectrum $\calp_\calr$ gives a
logarithmically divergent result for $\langle \calr^2 \rangle$,
but since $\calp_\calr\sim 10^{-9}$ one has to go to a huge
scale to see any effect. Taking the comoving small-scale cutoff
to be the Hubble distance at the end of inflation (which is equivalent
to the usual procedure of dropping the contribution of the
vacuum fluctuation to the energy density in  flat spacetime),
this  scale as a multiple of the Hubble distance is
\be
\frac{a_0  H_0}{k\sub{HU}}\sim\exp(10^{9}-60)\sim
\exp(10^{7})
\ee
If as expected the spectrum increases somewhat with scale this
estimate will be sharply reduced, but it will still be a big number
in typical models of inflation.

A possible divergence of the geometry distortion
associated with  a  nearly flat spectrum
 is interesting, because it
 suggests that the universe might be fractal on very large
scales \cite{eternal}. It cannot however be explored observationally,
because if the spectrum is fairly flat, large scales
give a negligible contribution
to the cmb anisotropy.
 Here we are concerned with the quite different
possibility, that the spectrum might rise sharply on some
not-so-huge  but still very large scale
 $q\sub{VL}$. We therefore suppose that the spectrum has the form
\be
\calp_{\calr}\su{VL}\simeq \delta(\ln k- \ln k\sub{VL}) \langle \calr^2
\rangle \label{vvlarge} \ee

Such a contribution
might originate from the vacuum fluctuation if the inflaton potential
has a suitable form, but more plausibly it would arise because the
vacuum assumption failed, or in other words because there were inflaton
particles with momentum $k\lsim k\sub{VL}$.

Before calculating its effect, let us spell out the physical significance
of this contribution. We will make the natural assumption that the homogeneous
Gaussian perturbation under consideration exists in a patch around us,
whose size is much bigger than the correlation length
\be
d\sub{VL}=a_0/k\sub{VL}
\label{dflat}
\ee
The curvature perturbation is more or less constant in a
region of size $d\sub{VL}$.
According to  Gauss-Bonnet theorem,  the
distortion in geometry in this region
(measured for instance by the departure of the
sum of the angles in a geodesic
triangle spanning it from $2\pi$)
is of order
its cross-sectional area times the perturbation in the curvature scalar.
{}From \eq{msqflat} this is of order
\be
d\sub{VL}^2\delta R^{(3)} \sim
\langle\calr^2\rangle\half
\ee
In some  patches  the curvature  perturbations  will
be positive, and in others  negative. It does not make sense
to consider a value $\langle\calr^2\rangle$ bigger than 1, because
then the regions of  positive curvature would close on themselves.

The geometry distortion in the observable universe is
of order $a_0^2\delta R^{(3)}$, and so smaller by a factor
$k\sub{VL}^2$.
As the correlation length increases
with fixed $\langle\calr^2\rangle$,
the geometry distortion in the observable universe decreases and so
does the spatial gradient   For both these reasons,
the anisotropy caused by a given value of $\calp_\calr$ decreases.
Let us calculate it.

Since $j_l(x)\sim x^l$ for small $x$, one sees
from \eq{clflat} that the quadrupole dominates.  Using $j_2(x)=x^2/15$ one
finds \be 6 C\su{VL}_2=\frac{4\pi}{25}\frac{16}{15^2}
 \rfrac{k\sub{VL}}{a_0H_0}^4 \langle
\calr^2 \rangle \label{27} \ee The quadrupole measured by COBE is not
significantly in excess of the typical values $l(l+1)C_l\simeq
8\times 10^{-10}$
of the other multipoles (in fact it is somewhat smaller), so we conclude that
$C_2\su{VL}$ is absent at this level.  This means that
\be
d\sub{VL}>70 \langle \calr^2 \rangle\quarter H_0\mone
 \label{gzbound} \ee

As Grishchuk and
Zeldovich pointed out, this bound on
$\langle \calr ^2\rangle$ becomes weaker as the scale increases.
It can be of order 1 provided that (cf. \cite{turner})
\be
\frac{a_0H_0}{k\sub{VL}}>70 \ee

A bound similar to this one  was already
 implied by upper limits
on the quadrupole that existed two decades ago.
But before COBE measured the actual values of the
low multipoles there was always the possibility that the Grishchuk-Zeldovich
effect might be present (ie., that the quadrupole might stick out
above the other multipoles), indicating a big curvature perturbation
on some very
large scale.  The  somewhat disappointing fact that the
 Grishchuk-Zeldovich is absent  seems not to
have been
noted anywhere in the copious literature on the COBE observations.

\subsection{The super-curvature scale Grishchuk-Zeldovich effect}

To generalize the Grishchuk-Zeldovich effect to
 $\Omega_0<1$ one needs to take spatial curvature into account,
and to note that the limit of large scales corresponds to
$k\to 0$, not $q\to 0$. This has not been done to date.
The only relevant publications of which we are aware are
 \cite{turner} where spatial curvature is ignored, and
\cite{kamsper,kashlinsky} where
the appropriate limit is incorrectly assumed to be $q\to 0$.
In contrast with the case $\Omega_0=1$,
the Grishchuk-Zeldovich effect for the case $\Omega_0<1$
could come only from a non-vacuum contribution.

Consider therefore a contribution on very large
super-curvature
scales, and represent it by
\be
\calp_\calr\su{VL}\simeq \delta(\ln k-\ln k\sub{VL})
\langle \calr^2 \rangle
\label{107}
\ee
We need to understand the physical significance of the corresponding
curvature perturbation, which is different from the case $\Omega_0=1$.
{}From Section 4.1, the correlation length is now
\be
d\sub{VL}=a_0/k\sub{VL}^2
\label{dcurved}
\ee
The curvature perturbation is practically constant in a region of this
size. The {\em perturbation} in the geometry distortion of such a
region is now
\be
d\sub{VL}^2 \delta R^{(3)}\sim k\sub{VL}\mfour \langle
\calr^2\rangle\half
\ee
On the other hand there is now a distortion even in the absence of
a perturbation, given by \eq{rthree},
\be
d\sub{VL}^2 R^{(3)}\sim k\sub{VL}\mfour
\ee
Thus $\langle \calr^2\rangle\half$ now measures the fractional
perturbation in the geometry distortion, not the distortion itself.
However the requirement that regions of space should not close on
themselves is still equivalent to
$\langle
\calr^2\rangle\half\lsim 1$.

An equivalent way of viewing $\langle
\calr^2\rangle\half$ is that it measures the geometry distortion
of a region with size $a_0$. Since $\Omega_0$ is not extremely small
this is the same as saying that it measures the geometry distortion
of the observable universe. As
the correlation length $d\sub{VL}$ is increased with $\langle
\calr^2\rangle$ constant, the geometry distortion
of the observable universe
does not decrease as it does in the $\Omega_0=1$ case.
One still expects, though, that for a given value of $\calp_\calr$
the effect on the cmb anisotropy will become
smaller,
because the spatial
gradients become smaller. Let us see how to calculate it.

As $k\to0$, $\Pi_{k0}\to 1$,
but the other radial functions are proportional to
$k$. The normalisation factor becomes
\bea
kN_{k1}&\to& N_1\equiv \sqrt{\frac2\pi} \\
 kN_{kl}&\to& N_l\equiv
\sqrt{\frac2\pi} [\prod_{n=2}^l(n^2-1)]\mhalf
\hspace{5mm}(l\geq 2)
\eea
and
\be
\tilde\Pi_{k1}(r)\to
\frac{k^2}{4}\frac1{\sinh^2 r}\left[\sinh(2r)-2r\right]
\ee
The other radial functions
follow from the recurrence relation
\eq{recur}. It is convenient to define
\be
\tilde\Pi_l\equiv\lim_{k\to0} \tilde\Pi\kl/k
\ee

Using these results, the  contribution to the  mean square
multipoles becomes
\be
C_l\su{VL} = N_l^2 B_l^2 k^{2}\sub{VL} \langle \calr ^2 \rangle
\label{gzlowom}
\ee
where
\be B_{l}\equiv\frac15 \tilde
\Pi_l(\eta_0)+\frac65\int^{\eta_0}_0 \tilde
\Pi_l(  r)F'(\eta_0-  r)\diff   r
\ee

When one increases the value of $l$ under consideration, the
scale above which these limits hold presumably becomes successively
larger, so to actually calculate $C_l\su{VL}$ for a given
value of $k\sub{VL}$ one ought to use the full expression
\eq{openswsc}, but hopefully \eq{gzlowom}
will provide a reasonable estimate for small $l$.
Since $N_l$ and $B_l$ are roughly of order 1 for low multipoles,
 it says very roughly that
\be
C_l\su{VL}\sim k^2\sub{VL} \langle\calr^2\rangle
\ee

The absence of $C_l\su{VL}$ at the level $10^{-10}$
therefore implies very roughly
\be
k\mtwo\sub{VL}\gsim 10^{10}\langle\calr^2\rangle
\label{gzopen}
\ee
Since $(a_0H)^2=1-\Omega_0$ is supposed not to be tiny, this  result
is roughly
\be
d\sub{VL}\gsim 10^{10}\langle\calr^2\rangle H_0\mone
\label{gzboundop}
\ee

In words, the conclusion is that
if the
 fractional geometry distortion
is of order 1, its correlation length must be more than
$10^{10}$ Hubble distances.
This result is not directly comparable with
 the $\Omega_0=1$ result, because it concerns
the fractional, not the
absolute,
geometry distortion on the scale $d\sub{VL}$.
 The quantity that measures the absolute geometry distortion
 is $\tilde\calr\klm\equiv k\sub{VL}^4
\calr\klm$, and in terms of this  quantity
\be
d\sub{VL} \gsim 100\langle  \tilde\calr^2 \rangle^{1/5}
H_0\mone
\ee
We see that if the absolute geometry distortion, in a region whose
diameter is equal to the correlation length, is of order 1, then
the correlation length  must be at least two orders of magnitude bigger
than the Hubble distance. This is essentially the same as the
$\Omega_0=1$ result.

These  estimates have been derived from the fact that the
Grishchuk-Zeldovich effect cannot be bigger than the
observed values of the multipoles. As we have not
actually calculated the $l$ dependence of $C_l\su{VL}$ we
cannot say that the effect is definitely absent  as in
the $\Omega_0=1$ case, because it might turn out that
the dependence of $C\sub{VL}$ mimics the dependence
that of the data (roughly $C_l\propto l\mtwo$).
It would be desirable to calculate the shape of $C_l\su{VL}$,
both to check that this does not happen and  to check the assumption
that $B_l$ is of order 1.

In contrast with the case $\Omega_0=1$, the Grishchuk-Zeldovich  effect
is present in all of the low multipoles if $\Omega_0<1$. Indeed,
it could even occur in multipoles $l\gsim 10$ in which case the
Sachs-Wolfe approximation would become inadequate to investigate it
and a full calculation would be necessary. The necessary formalism to
perform such a calculation is already in place \cite{gouda}, and
it has already been used for the sub-curvature modes
\cite{both,kamspersug,gorskiopen}. The extension to the
super-curvature modes raises no new issue of principle.

\subsection{The  physical significance of the
Grishchuk-Zeldovich effect}

In some of the literature
\cite{turner,kashlinsky},
 the absence of the Grishchuk-Zeldovich
effect at a given level has been regarded as evidence that
the smooth patch of the universe which we occupy
extends beyond the edge of the observable universe.
As we now explain, this is not the case.

To make the simplest point first, it is clear from \eqs{73}{74}
that the multipoles $a_{lm}$ of the
 cmb anisotropy depends only on the curvature perturbation
within the observable universe. This remains true when we
consider their ensemble mean squares $C_l$. Strictly interpreted,
the Grishchuk-Zeldovich effect just explores the effect of very small
spatial gradients of the curvature perturbation, within the observable
universe, on the hypothesis that the curvature perturbation is a
typical realization of a homogeneous Gaussian random field.
Recall that in this context
`homogeneous' means that the correlation function of the curvature
perturbation depends only on the distance between two points, not on
their location; but we are still talking about locations within the
observable universe.

If this hypothesis is indeed correct within the observable universe, one
expects it to remain correct in some larger region. {\em If} this region
is sufficiently big, one can introduce the concept of a correlation
length as we did in the above discussion. By definition, the correlation
function is more or less constant out to a distance of order the
correlation length, only then falling off. Clearly `sufficiently
 big' means bigger than the correlation length.
But we will never know whether this picture is correct,
because we will never know what lies beyond
the edge of the observable universe (except by waiting for it to gradually
recede, in comoving distance units).

Homogeneity of the perturbation corresponds to the spectrum
(defined by \eq{fullspec}) being independent of $l$ and $m$.
One can reasonably expect this property to
fail when $k$ becomes so small that the corresponding
distance $a_0/k\sub{VL}^2$ becomes bigger than the size of the
smooth patch of the universe around us,
within which the perturbation is homogeneous.
In that case the {\em presence} of the Grishchuk-Zeldovich
effect on a given scale would suggest, though not really prove,
that the smooth patch extends to the corresponding distance. But
its absence says nothing.

Beyond the smooth patch might be regions of the universe
where the `perturbations' become so big that it makes no sense to talk
about a Robertson-Walker universe. If so the patch discussed
in the last paragraph will have a periphery, within which the typical
magnitude of the perturbations becomes bigger as one moves outwards
(in contrast with the region within the patch, where
the typical magnitude is by definition the same everywhere).
The Grishchuk-Zeldovich effect tells us absolutely nothing
about this periphery.

For the case $\Omega_0=1$ one might
argue that information about the periphery is available, by taking
the density perturbation to be the primary quantity rather than
the curvature perturbation.
{}From this viewpoint the large density perturbation in the periphery
will generate a large curvature perturbation in the observable universe,
through the usual `Coulomb law' solution of the Poisson equation
\eq{densflat}, unless there is an accidental cancellation.
 However there does not seem to be any
justification for it, and it does not work for
$\Omega_0<1$ because according to \eq{density}
the Poisson equation does not hold. Rather, the
density perturbation and the curvature
perturbation become essentially the same on scales much bigger than
the curvature scale.

\subsubsection*{Inflation and the Grishchuk-Zeldovich effect}

A separate issue is whether the absence of the Grishchuk-Zeldovich
effect tells us anything about inflation. This is clearly the case
only if the scale $k\sub{VL}$ can be related to inflation.
Reference \cite{turner}, which deals with the case $\Omega_0=1$
(or at any rate ignores spatial curvature), accepts the usual dogma that
`the universe is smooth on some scale of order the Hubble distance
at the beginning inflation'. Interpreting this to mean
that $k\sub{VL}\sim aH$ at the beginning of exponential
($\Omega=1$) inflation, the absence of
the Grishchuk-Zeldovich
effect indeed tells us that inflation starts several Hubble times
before the observable universe leaves the horizon.
However, as discussed in Section 5.3
the usual dogma does not have any
clear justification.

\section{Conclusion}

In this article we have drawn the attention of physicists to the
incompleteness of the standard mode expansion for cosmological
perturbations in an $\Omega_0<1$ universe. In order to generate the
most general homogeneous Gaussian random field one should
use \eq{fullspec}, which runs over all negative eigenvalues
$-k^2$,
whereas the standard expansion keeps only the modes with
$k^2>1$. We have called these sub-curvature modes, because
they vary appreciably over a distance less than the curvature scale,
and we have called the modes with $0<k^2<1$ super-curvature modes.

The fact that super-curvature modes are needed to generate the most
general perturbation has been known to mathematicians for about half a
century, so that their omission by cosmologists
constitutes a remarkable failure of communication between the
worlds of mathematics and science. This omission leads to
perturbations which are practically uncorrelated on scales bigger than
the curvature scale. In contrast, a mode with $k^2\ll 1$
corresponds to a perturbation with correlation length
$k\mtwo$ in units of the curvature scale.

What nature has chosen to do is of course another question.
For the case $\Omega_0=1$ the standard assumption is that the
perturbations originate as the vacuum fluctuation of the inflaton field,
and in 1990 this assumption was extended to the case
$\Omega_0<1$ \cite{lystomega}. The mode expansion of a quantum
field runs only over sub-curvature modes, since they form a complete
orthonormal set for square integrable function. As a result the
vacuum fluctuation generates a perturbation which includes only these
modes.

Of course this is not a feature of quantum field theory
{\em per se}, but of the assumption that the inflaton field is in
the vacuum. This assumption might break down below some
small value of $k$, corresponding to a correlation length much bigger
than the curvature scale. We therefore ask whether
a big perturbation with a very large correlation length could
be detected through
the cmb anisotropy. For the case $\Omega_0=1$, this
question was asked and answered in 1978 by Grishchuk and Zeldovich
\cite{grze}, and here we have extended their discussion to the case
$\Omega_0<1$ by including the super-curvature modes. We have given
a formula for the cmb anisotropy due to a mode with $k\ll 1$,
and have estimated its
magnitude. By requiring that it be no bigger than the observed
anisotropy we have estimated a lower limit on the correlation length
for a perturbation of given magnitude. As in the case
$\Omega_0=1$, the correlation length must be more than about two
orders of magnitude bigger than the size of the observable universe,
if the geometry distortion is of order 1  in a region whose size
is equal to the correlation length.

In contrast with
the case $\Omega_0=1$, the Grishchuk-Zeldovich effect is present
in all multipoles, not just in the quadrupole. It would be interesting
to evaluate its $l$ dependence, if only to check that it does not
mimic the observed dependence which is usually interpreted as coming
entirely from the vacuum fluctuation.

\section*{Appendix}

This appendix   gives some mathematical results, in the sort of language
that we as physicists are accustomed to. It deals with both  the
spherical expansion used in the text, and with  an expansion using coordinates
that  slice space into flat  surfaces which  is more like the
flat-space Fourier series.
 At the risk of being pedantic we  give
a rather full treatment, because even in the flat-space case
there does not seem to be a  reference that explains the basic concepts
in a way that is accessible to most physicists.

We refer the reader seeking a rigorous but more abstract treatment
to  \cite{yaglom}.

\subsection*{The Fourier expansion}

In flat space the simplest approach is to use  the Fourier expansion.
In comoving coordinates $\bfr\equiv (x,y,z)$
it is
 \be
f(\bfr)=(2\pi)\mthreehalf\int \diff^3\bfq f\sq
 \exp(i\bfq.\bfr) \label{four}
\ee
The orthonormality relation is
\be
(2\pi)\mthree \int \diff {\cal V} \exp(-i\bfq.\bfr)  \exp(i\bfq'.\bfr)
=\delta^3(\bfq'-\bfq)
\label{orthofou}
\ee
where $\diff{\cal V}=\diff^3\bfr$ is the volume element.

 A Gaussian perturbation
is obtained by assigning independent Gaussian probability distributions
to the  real and imaginary parts of the coefficients $f\sq$, but
we need to ensure translation and rotation invariance of the
correlation function. Translation invariance is equivalent to
the real and imaginary parts having the same distributions
(same mean squares), for the following reason.
Because $\exp[i(\bfq.\bfx)]\exp[i(\bfq'.\bfx')]$
is a function of $\bfx-\bfx'$ only if $\bfq=-\bfq'$, translation
invariance is equivalent to introducing a correlation between
$f\sq$ and $f\msq$ only, which because of the reality condition
 $f^*\sq=f\msq$ means a correlation between
$f\sq$ and $f^*\sq$ only. This means that the phases of
$f\sq$ must be uncorrelated, which indeed means that the real and imaginary
parts of $f\sq$ must have the same distribution.\footnote
{A direct way of seeing this is to work with the real form of the Fourier
integral and note the identity $\cos a\cos b+\sin a \sin b=\cos(a-b)$}
Let us therefore define the spectrum by
\be
\langle f\sq^*
 f\sqp \rangle
 =\frac{2\pi^2}{q^3} \calp_f(q)
\delta(\bfq-\bfq') \label{specfou} \ee
The correlation functions is then
\be
\xi_f(r)=(2\pi)\mthree 2\pi^2\int \frac{\diff^3\bfq}{ q^3} \exp(-i\bfq .\bfr
) \calp_f(q) \label{12} \ee
Rotational invariance is clearly
 equivalent to the spectrum depending only on the
magnitude of $\bfq$, not its direction.
 Performing the angular integration one obtains \eq{corrflat}.

Using the well known expansion of a plane wave into  spherical
waves one can prove that  the above definition of the
spectrum is equivalent to the definition \eq{flatspec}
in terms of the spherical expansion. The equivalence, and in particular
the fact that the two definitions are the same except for the different
delta functions, does not depend on the detailed form of
 the transformation
between the spherical expansion and the Fourier expansion,
but rather on the fact that it  is unitary.

\subsection*{The spherical  expansion}

We first justify the claim made in the text, that the
correlation function depends only on the distance between the points
if the spectrum defined by \eq{29} is independent of
$l$ and $m$.
 Let
$  r,\theta,\phi$ and $  r',\theta',\phi'$ be the coordinates of a given
point with respect to two different spherical coordinate systems.
We saw earlier that the most general eigenfunction
with eigenvalue  $-(k/a)^2$
is a linear combination of the functions
$Z_{klm}(  r,\theta,\phi)$.  This is of course true in any
coordinate system.  Since $Z\klm(  r,\theta,\phi)$ and
$Z\klm(  r',\theta',\phi')$ are both eigenfunctions
it follows that either of them can be expanded in terms of the other.
Thus, there is a linear combination of the form
\be
Z\klm(  r,\theta,\phi)=\sum_{l'm'} U^k_{lml'm'}
Z_{kl'm'}(  r',\theta',\phi') \label{unitary} \ee

If the spectrum defined by \eqs{29}{29a}
is independent of $l$ and $m$, the correlation function given by
\eq{corr4} or its super-curvature analogue
will be invariant under transformations of the above form
provided that the transformation matrix satisfies the unitarity
property
\be
\sum\sub{l''m''} U^k_{lml''m''} (U^k_{l'm'l''m''})^*=\delta_{ll'}\delta_{mm'}
\ee
Since the transformation takes a pair of points into arbitrary
positions subject to the constraint that the distance between them is
fixed, the correlation function will then depend only on this
distance.

For the sub-curvature modes, unitarity
follows from the fact that the
transformation takes one orthonormal basis (for the subspace of
eigenfunctions with a given eigenvalue) into another.  Let us see this
explicitly.  Orthonormality for the whole $L^2$ space gives the the
coefficients of the expansion as \be \delta(q'-q) U^k_{lml'm'}= \int
Z^*_{k'l'm'}(  r',\theta',\phi') Z_{klm}(  r,\theta,\phi) \diff{\cal V}
\label{uexpress} \ee where the primed coordinates are regarded as functions
of the unprimed ones and the volume element is defined by \eq{volume}.  Now
consider the inverse transformation, \be
Z_{kl'm'}(  r',\theta',\phi')=\sum_{lm}
V^k_{l'm'lm}Z_{klm}(  r,\theta,\phi) \ee The coefficients are given by \be
\delta(q'-q) V^k_{l'm'lm}=\int
Z^*_{k'lm}(  r,\theta,\phi)Z_{kl'm'}(  r',\theta',\phi') \diff{\cal V}'
\label{vexpress} \ee where now the unprimed coordinates are regarded as
functions of the primed ones.  But as the integration goes over all space one
can just as well integrate over the unprimed coordinates and regarded the
primed coordinates as the dependent ones.  By comparing
\eqs{uexpress}{vexpress} it follows that the transformation is indeed
unitary,
\be \left(U^k_{lml'm'}\right)^*=V^k_{l'm'lm} \label{unidef} \ee

This proof of unitarity  does not work for the
super-curvature regime, because
we invoked orthonormality to obtain
\eq{uexpress} for the matrix element
$U^k_{lml'm'}$.  There is however an alternative expression that remains well
behaved in the super-curvature regime,
 obtained by substituting into \eq{unitary} the definition
of the $Z$'s, and remembering that the spherical harmonics $Y_{lm}$ are a
complete orthonormal set on the sphere.  Choosing {\em any} sphere
$  r'=$constant this expression is
\be \Pi_{kl'}(  r') U^k_{lml'm'} =
\int \Pi_{kl}(  r) Y_{lm}(\theta,\phi) Y^*_{l'm'}(\theta',\phi')
\sin\theta'\diff\theta'\diff\phi'
\label{uexpress2} \ee
where the unprimed
coordinates are regarded as functions of the primed ones.
(For sub-horizon
modes the original expression \eq{uexpress} is recovered if we multiply both
sides of \eq{uexpress2} by $\Pi_{k'l'}(  r')\sinh^2  r' \diff   r'$ and
integrate over $0<  r'<\infty$, but continuing to imaginary $q$
and $q'$ causes the
integrals on both sides to diverge at $  r'=\infty$.)  Similarly, choosing
any sphere $  r=$constant one has
\be
\Pi_{kl}(  r) V^k_{l'm'lm}  = \int
\Pi_{kl'}(  r') Y_{l'm'}(\theta',\phi') Y^*_{lm}(\theta\phi)
\sin\theta\diff\theta\diff\phi
\ee
In contrast with the original expressions
\eqs{uexpress}{vexpress}, the radial coordinate on the right hand side of
these new expressions runs only over a finite range.
When we analytically continue to imaginary $q$, the radial functions
pick up a factor $i$ which cancels, so the
equality $\left(U^k_{lml'm'}\right)^*=V^k_
{l'm'lm}$ that we established for real $q$
remains valid. Note that this does not  work for  complex eigenvalues,
indicating that they are not allowed.

The above discussion is a generalisation of the familar
demonstration  that a
rotation around the origin acts on $Y_{lm}$ with a {\em finite dimensional}
unitary matrix acting on the $m$ index alone.
  Because of the finite
dimensionality, this provides a {\em rigorous} proof that invariance under
such rotations is equivalent to the independence of the spectrum on $m$.  The
extension to arbitrary rotations and translations involves an infinite sum
over $l$,
and as we have not discussed its convergence the discussion is
not rigorous. Sasaki and Tanaka \cite{misaolatest}
have recently demonstrated that
the infinite sum over $l$ that occurs in
\eq{corr4} is uniformly convergent, which makes
the above derivation rigorous.

\subsection*{The flat-surface expansion}

The flat-surface expansion uses    coordinates
 defined by the line
 element \cite{fock,vilenkin,wilson}
 \be \diff
 l^2=(a/z)^2(\diff x^2 +\diff y^2 + \diff z^2)
 \ee
 The surfaces  of constant $z$ are flat. Any point in space can be chosen
 as the point $x=y=0$, $z=1$ and curvature is negligible in the region
 around that point $|x|\ll1$, $|y|\ll1$, $|z-1|\ll1$.

 Since a sphere of infinite radius is  flat, one can
 think of the flat surfaces $z=$constant as spherical wave fronts,
 originating from a point at $z=0$. Note that these surfaces
 have an `inside' and an `outside' even in the limit where they
 are flat, because geodesic surfaces such as
the `equatorial plane' $\phi=\pi/2$ are
 {\em not}  flat.

The virtue of these coordinates is that the form of the line element is
invariant under the following transformation
\bea
x&\to& C (x+X) \nonumber\\
y&\to& C (y+Y) \nonumber\\
z&\to& C  z \label{trans}
\eea
With a suitable choice of the constants $X$, $Y$ and $C$
we can place one of the two points to which the correlation function
refers at an arbitrary position,
while leaving unchanged the form of the  line element and
therefore the geodesic distance to the other point.

If the point $x',y',z'$
corresponds to the point $  r=0$ in the spherical coordinate system,
one can orientate the axes so that \cite{wilson}
\bea
x-x' &=& z'\cos\phi\sinh  r/\cosh(\eta-  r)\nonumber\\ y-y' &=&
z'\sin\phi\sinh  r/\cosh(\eta-  r)\nonumber
\\ z &=& z'\cosh\eta/\cosh(\eta-  r)
\label{coords}
\eea
where
  $\tanh\eta\equiv\cos\theta$.

Now consider the mode expansion.
	We look for eigenfunctions of the
form
\be W_{k\sqperp}= \int \diff\qperp^2 F_{k\sqperp}(z)e^{i\sqperp.\sbfx}
\ee
where $\bfx$ is the vector with components $x,y$.
 Substituting this expression into the Laplacian
gives a second order equation for $F_{k\sqperp}$. One of its two
 linearly independent solutions is \cite{vilenkin,wilson}
 $zK_{iq}(q_\perp z)$ where
$K$ is the modified Bessel function.
 This solution vanishes at $z=\infty$, and
up to normalisation is the only solution with that
property. It  also
vanishes at $z=0$ (spatial infinity in opposite direction)
for real $k$.
Note that $K_{iq}$ is real for  real $q^2$ and imaginary for
imaginary $q$.

We will
define the sub-curvature modes by
\be
W_{k\sqperp}(\bfx,z) \equiv N(k)
 e^{i\sqperp.\sbfx} z K_{iq}(q_\perp z)
\ee
where
\be
N^2(k)= \frac{q\sinh(\pi q)}{2\pi^4}
\ee
The normalisation  has been chosen to satisfy the orthonormality condition
\be
\int W_{k\sqperp}(\bfx,z) W_{k'\sqperp'}(\bfx,z) \diff{\cal V}
=\delta^2(\qperp-\qperp')\delta(q-q')
\ee
where the volume element is
$\diff{\cal V}=\diff x \diff y \diff z/z^3$.
This condition is equivalent to
 the orthonormality of  the functions $(2\pi)\mone  e^{i\sqperp
.\sbfx}$ plus the relation
\cite{lebedev,erdelyi,erdelyi2}
\be
\frac{2\pi\mtwo} q\sinh  (\pi q)\int^\infty_0 K_{iq}(z)
K_{iq'}(z) \diff z/z =\delta(q-q')
\ee
 The expansion of a generic perturbation \cite{lebedev}
 in terms
of these functions is
\be
f(\bfx,z)=\int_0^\infty \diff q \int \diff^2\qperp
 f_{k\sqperp} W_{k\sqperp}(\bfx,z)
\label{wexpansion}
\ee

Following Wilson \cite{wilson},
we construct a Gaussian perturbation by assigning independent Gaussian
distributions to the real and imaginary parts of the coefficients.
Translation invariance in the $\bfx$ plane requires that the real
and imaginary part of each coefficient has the same mean square,
so we define the spectrum by (cf.~\eq{29})
\be
 \langle f^*_{k\sqperp} f_{k'\sqperp'} \rangle =
\frac{2\pi^2}{q(q^2+1)} \calp_f(k)
\delta(q-q')\delta^2(\qperp-\qperp')
\label{appspec} \ee
  The correlation function is
\be \xi_f(\bfx,z,\bfx',z')= \int^\infty_0 \diff q
\frac{2\pi^2N^2(k)}{q(q^2+1)}
 \calp_f(k) \int\diff^2\qperp
e^{i\sqperp.(\sbfx-\sbfx')}
zK_{iq}(q_\perp z)z' K_{iq}(q_\perp z') \ee
It is invariant under \eq{trans},
because the factor $C$ appearing in \eq{trans}
can be absorbed into the definition of the integration variable
$\qperp$.

 Using \eq{coords} and integrating over the angular direction
in the $\qperp$ plane, the correlation function becomes
\be \xi_f = 2\pi \beta
\int^\infty_0 \diff q \frac{\sinh (\pi q)}{\pi^2(q^2+1)} \calp_f(k)
\int^\infty_0 \diff p pJ_0(p\gamma) K_{iq}(p) K_{iq}(\beta p)
\label{ancorr}
\ee
where
\bea \beta &\equiv& \frac {\cosh
\eta}{\cosh(\eta-  r)} \label{zeq}\\ \gamma &\equiv& \frac {\sinh
  r}{\cosh(\eta-  r)} \label{xeq} \eea
 From Eq.~(8.13.30) of \cite{erdelyi}, \be \int^\infty_0 \diff p p J_0(
p\gamma) K_{iq}(p) K_{iq}(\beta p) =\frac{ \sqrt\pi \Gamma(1+iq)\Gamma(1-iq)}
{ 2\twothird \beta(u^2-1)\quarter } P^{-1/2}_{iq-1/2}(u) \label{an5} \ee
where
\be u\equiv \frac{\gamma^2+\beta^2+1}{2\beta} =\cosh  r\ee
This is a function only of $  r$, and using \eq{pileg} one finds
that the correlation function is given by
 \eq{corr2}.

Now consider the super-curvature modes $-1<q^2<0$. The normalisation
factor $N(k)$ becomes purely imaginary so it is convenient to drop
the $i$ factor, defining
\be
N^2(k)=\frac{|q|\sin(\pi  |q|)}{2\pi^4}
\ee
The super-curvature contributions are
\bea
f\su{SC}(\bfx,z)&=&\int_0^1 \diff (iq) \int \diff^2\qperp
 f_{k\sqperp} W_{k\sqperp}(\bfx,z)
\label{wexpansionsc}\\
 \langle f^*_{k\sqperp} f_{k'\sqperp'} \rangle &=&
\frac{2\pi^2}{|q|(q^2+1)} \calp_f(k)
\delta(iq-iq')\delta^2(\qperp-\qperp')
\label{appspecsc}
\eea
The correlation function is given by \eq{ancorr} with $q\to iq$,
and since \eq{an5} is valid for $-1<iq<1$  this proves \eq{corr3}.

The spherical and flat-surface expansions are equivalent, at least
in the present context, because they both lead to  a Gaussian
perturbation with the  same correlation function.

\subsection*{The flat-space limit of the flat-surface expansion}

We end by looking at  the flat-space limit of
the flat-surface   expansion. Though not strictly necessary for our purpose,
it is extremely  instructive, and
does not seem to have been given before.

The coordinates $x$, $y$ and $z$ become Cartesian in a
a small region around $x=y=0$ and $z=1$, and
then  $q_x$, $q_y$ and $q_z\equiv \sqrt{q^2-q_x^2-q_y^2}$
are the would-be components of the  vector in the Fourier
expansion. But the expansion \eq{wexpansion}  does not  restrict the
range of $q_x$ and $q_y$, so it will include both real and imaginary
$q_z$. In other words it will include hyperbolic  functions as well
 as circular ones.  Evaluating the limiting behaviour of $K_{iq}$
confirms this. One finds \cite{erdelyi2}
for real $q_z$
\be
K_{iq}(q_\perp z)\to A \sin(B +q_z \tilde z)
\ee
where $\tilde z\equiv z-1$,
 $A\equiv\sqrt{2\pi} q_z\mhalf e^{-\pi q/2}$
and $B\equiv\frac\pi4+q\cosh\mone(q/q_\perp)-q_z$.
For imaginary $q_z$ one finds
\be
K_{iq}(q_\perp z)\to C \exp(D-|q_z|\tilde z)
\ee
where $C\equiv(2|q_z|/\pi)\half$ and
$D=-q_\perp \sin\mone (q/|q_z|)$.

The second  expression becomes  infinite when $|q_z|
\tilde z\to -\infty$, but even so
 a translation invariant correlation function
will result  when  the two expressions  are substituted into
 \eq{ancorr}.
This example  serves to remind us that we should take nothing
for granted
when considering which modes are allowable.

\subsection*{Acknowledgements}

This work was started with the help of EU research grant
ERB3519PL920782(10835).
 One of us (DHL) thanks the Isaac Newton Institute for a visiting
Fellowship while the work was being completed,
and    Bruce
Allen,  Robert Caldwell,
 Misao Sasaki and Neil Turok
for useful discussions there and for communicating in advance  their results.
  We also  thank Andrew Liddle for
useful  comments on an earlier version of the draft.



\footnotesize
\begin{thebibliography}{99}

\bibitem{colesellis} P.  Coles and G.  Ellis,
 Nature  {\bf 370}, 609 (1994).
  \bibitem{dekelrev} A.  Dekel,  Ann.  Rev.
Astron.  Astroph., {\bf 32}, 371 (1994).
\bibitem{loomega} G.  F. R. Ellis, D. H. Lyth
and M. B. Mijic,  Phys. Lett B {\bf   271}, 52 (1991).
 \bibitem{lystomega} D.  H.  Lyth and E.  D.
Stewart, Phys Lett. B {\bf 252}, 336 (1993).
 \bibitem{smet} G.  F.  Smoot {\it et.
al.}, {\it Astrophys.  J.  Lett.}  {\bf 396} (1992) L1.
\bibitem{gorski} K.  G\'{o}rski {\em et al.},
Astroph. J. Lett.  {\bf 430}, 89 (1994).
  \bibitem{bunn} E. F. Bunn, D. Scott and M. White,  preprint
 astro-ph/9409003  (1994).
 \bibitem{fock} V. Fock, Z. Physik {\bf 98}, 148 (1935).
\bibitem{bander}  M. Bander and C. Itzykson, Rev. Mod. Phys.
{\bf 38} 346
(1966).
\bibitem{grze} L. P. Grishchuk  and Ya. B. Zel\'{}dovich,
Astron. Zh.  {\bf 55}, 209 (1978) [Sov. Astron. {\bf 22}, 125 (1978)].
\bibitem{yaglom}
A. M. Yaglom, in {\sl Proceedings of the Fourth Berkeley Symposium
Volume II}, edited by  J. Neyman (University of California Press, Berkeley,
1961).
\bibitem{krein} M. G. Krein, Ukrain. Mat. Z. {\bf 1}, No. 1, 64 (1949);
{\em ibid} {\bf 2}, No. 1, 10 (1950).
\bibitem{lifs} E.  M.  Lifshitz, J.  Phys.  (Moscow) {\bf 10},
116 (1946);
E. M. Lifshitz and I. M. Khalatinikov, Adv.  Phys. {\bf 12},
 185 (1963).
  \bibitem{wilson} M. L. Wilson, Astrophys. J.  {\bf 273}, 2 (1983).
\bibitem{kotu} E. W. Kolb and M.  S.  Turner,
{\em The Early Universe} (Addison-Wesley 1990).
\bibitem{dolgov}  A. Z. Dolginov and I. N. Toptygin,
Soviet Physics JETP,  {\bf 37(10)}, 1022 (1960).
\bibitem{vilenkin} N. Ya. Vilenkin and Ya. A. Smorodinsky,
Soviet Phys. JETP, {\bf 19}, 1209 (1964).
\bibitem{harrison}
E. R. Harrison, Rev. Mod. Phys.  {\bf 39}, 862 (1967).
\bibitem{bida} N. D. Birrel and P. C. W. Davies,
{\em Quantum Field Theory in Curved Space-Time}
(Cambridge University Press 1982).
\bibitem{fabbri} R. Fabbri, I. Guidi and V. Natale,  Astrophys.
J. {\bf 257} 17 (1982).
\bibitem{gelfand} I. M. Gelfand, M.
I. Graev and N Ya Vilenkin, {\sl Generalized Functions: Volume 5; Integral
Geometry and Representation Theory} (Academic Press Inc., New York,
1966);
M. A.  Naimark, {\em Linear Representations of the Lorentz Group}
(Pergamon Press 1964).
 \bibitem{adler} R. J. Adler, {\em The Geometry of Random
Fields} (Wiley, Chichester,  1981).
\bibitem{peebles} P.  J.  E.  Peebles, {\sl The Large Scale Structure of the
Universe } (Princeton University Press, 1980).
\bibitem{chaotic} A.  D.
Linde, {\it Phys Lett} {\bf B129}, 177 (1983).
\bibitem{bubble} S. Coleman and F. De Luccia,  Phys. Rev.
D  {\bf 21},  3305 (1980); J. R. Gott, Nature {\bf 295} , 304
(1982);  A. H. Guth and E. J. Weinberg, Nucl. Phys. {\bf B212},
321 (1983);
 J. R. Gott and T. S. Statler, Phys. Lett. {\bf B136},
157 (1984).
\bibitem{neil}
M. Bucher, A. Goldhaber and N. Turok, preprint (1994).
\bibitem{misaonew}
 M. Sasaki, T. Tanaka, K. Yamamoto and J. Yokoyama,
Phys. Lett. B  {\bf 317}, 510 (1993);
M. Sasaki, T. Tanaka, K. Yamamoto and J. Yokoyama,
Prog. Theor. Phys. {\bf 90}, 1019 (1993);
T. Tanaka and M. Sasaki, preprint (1994). T. Tanaka and
 M. Sasaki, Phys. Rev. D , in press   (1994);  K. Yamamoto,
T. Tanaka and M. Sasaki, preprint (1994).
\bibitem{neilnew} M. Bucher and  N. Turok, in preparation.
\bibitem{karlin} S. Karlin and H. M. Taylor, {\sl A first course
on stochastic processes} (Academic Press, New York, 1975).
\bibitem{bbks} J.  M.  Bardeen, J.  R.  Bond, N.  Kaiser and A.  S.  Szalay,
{\it Astrophys. J.} {\bf 304}, 15 (1986).
\bibitem{wilsonfirst} M. L. Wilson, Astrophys. J. Lett.,
{\bf 253}, 53 (1982).
\bibitem{early} K. M. Gorski and J. Silk, Astrophys. J. {\bf 346},
L1 (1989).
 \bibitem{suggou} N.
Sugiyama, N.  Gouda and M.  Sasaki, Astrophys.  J.  {\bf 365}, 432 (1990).
\bibitem{gouda} N. Gouda, N. Sugiyama and M.
Sasaki, Prog. Theor.
Phys., {\bf 85}, 1023 (1991).
\bibitem{kamsper} M.  Kamionkowski and D.  N Spergel,
 Astrophys.  J. {\bf 432}, 7 (1994).
 \bibitem{both} N.  Sugiyama and
J.  Silk, Phys. Rev. Lett. {\bf 73}, 509 (1994).
\bibitem{kamspersug} M.
Kamionkowski, D.  N.  Spergel and N.  Sugiyama,  Astrophys.  J.  {\bf
426} (1994) L57.
  \bibitem{kametal} M.  Kamionkowski, B.  Ratra, D.  N.
Spergel and N.  Sugiyama, ``CBR anisotropy in an open inflation, CDM
cosmogony'', Princeton preprint (1994).
 \bibitem{ratrapeeb} B.  Ratra and
P.  J.  E.  Peebles, Astroph. J., {\bf 432}, 5 (1994);
B.  Ratra and
P.  J.  E.  Peebles, preprint (1994).
  \bibitem{tegsilkopen} M.  Tegmark and J.  Silk,
``Reionization in an open CDM universe:  implications for cosmic microwave
background fluctuations'', Berkeley preprint (1994).
  \bibitem{kashlinsky}
A.  Kashlinsky, I.   Tkachev and J.  Friedman,
Phys. Rev. Lett., {\bf 73}, 1582 (1994).
  \bibitem{bardeen} J.  M.  Bardeen,
Phys.  Rev.  D, {\bf 22}, 1882 (1980).
  \bibitem{ly85} D.  H.  Lyth, Phys.  Rev.  D {\bf 31}, 1792 (1985).
  \bibitem{lymu} D.  H.  Lyth and M.
Mukherjee,  Phys.  Rev.  D {\bf 38}, 485 (1988);
 D.  H.  Lyth and E.  D.
Stewart,  Astrophys.  J.  {\bf 361}, 343 (1990).
  \bibitem{ellis}
G.  F.  R.  Ellis and M.  Bruni,  Phys.  Rev.  D  {\bf 40}, 1804 (1989);
 M.  Bruni, P.  K.  S.  Dunsby and G.  F.  R.  Ellis,  Astrophys.
J.  {\bf 395}, 34 (1992); P.  K.  S.  Dunsby, M.  Bruni and G.  F.  R.
Ellis,  Astrophys.  J.  {\bf 395}, 54 (1992).
\bibitem{LL2} A.  R.  Liddle and D.  H.  Lyth,  Phys.  Rep.  {\bf 231},
1 (1993).
\bibitem{ll94} A. R. Liddle and D. H. Lyth, Mon. Not. R. Astron.
Soc., to be published (1994);
A. R. Liddle and D. H. Lyth, in preparation.
\bibitem{ks84} H.  Kodama and M.  Sasaki,  Prog.  Theor.  Phys.  {\bf
78}, 1 (1984).
 \bibitem{mukrev} V.F.  Mukhanov, H.  A.
Feldman and R.  H.  Brandenberger,  Phys.  Rep.  {\bf 215}, 203 (1992).
\bibitem{sb89} D. S. Salopek, J. R. Bond and J. M. Bardeen, Phys. Rev.
{\bf D40}, 1753 (1989).
\bibitem{bs83}  J.  M.
Bardeen, P.  S.  Steinhardt and M.  S.  Turner,  Phys.  Rev.  D {\bf 28},
679 (1983).
\bibitem{new} D. H. Lyth and E. D. Stewart, in preparation;
M. Bruni and P. K. Dunsby, Int. Journ. Mod. Phys
{\bf 3}, 443 (1994).
\bibitem{lyst90} D. H. Lyth and E. D. Stewart, Astrophys. Journ.
{\bf 361}, 343 (1990).
\bibitem{brly} M. Bruni and D. H. Lyth, Phys. Lett. B {\bf 323},
118 (1994).
\bibitem{adpred} A.  A.  Starobinsky,  Phys.  Lett.  {\bf B117},
175 (1982);
 S.  W.  Hawking,  Phys.  Lett.  {\bf B115}, 339 (1982);
 A.  H.  Guth
and S.-Y.  Pi,  Phys.  Rev.  Lett.  {\bf 49}, 1110 (1982).
  \bibitem{mukhanov} V.  F.  Mukhanov,  JETP Lett.  {\bf 41}, 493
(1985).
\bibitem{sasaki} M.  Sasaki,  Prog.  Theor.  Phys.  {\bf
76}, 1086 (1986).
Phys. Lett. B {\bf 323}, 118 (1994).
\bibitem{edet} E. J. Copeland, A. R. Liddle,  D. H. Lyth, E.
D. Stewart and D. Wands,   Phys. Rev. D
{\bf 49}, (1993), page 6427.
\bibitem{lyth84} D. H. Lyth, Phys Letts {\bf 147B}, 403 (1984).
\bibitem{halliwell} J. J. Halliwell and S. W. Hawking,
Phys. Rev. D {\bf 31}, 1777  (1985).
\bibitem{eternal} A. D. Linde,
  Mod. Phys. Lett, {\bf A1}, 81 (1986); A. D. Linde,
{\sl Particle Physics and Inflationary Cosmology}
(Harwood, Chur, Switzerland, 1990).
\bibitem{linde} A. D. Linde, `Lectures on Inflationary Cosmology'
unpublished (1994).
\bibitem{ewan} E. D. Stewart, personal communication.
\bibitem{ewanstuff} E. D. Stewart, preprint hep-ph/940539, to be
published in Phys. Rev. D; E. D. Stewart,
preprint astro-ph/9407040, to be published in Phys. Letts. B;
E. D. Stewart, preprint hep-ph/9408302;
D. H. Lyth and E. D. Stewart, preprint hep-ph/9408342,
to be published in the proceedings of
`Birth of the Universe and Fundamental Physics';
 G. Dvali, Q. Shafi and R. Schaefer, preprint hep-ph/9406139 (1994).
\bibitem{stewly}
E. D. Stewart and D. H. Lyth, Phys. Lett. B {\bf 302}, 171 (1993).
\bibitem{starobinsky} A. A. Starobinsky, Sov. Astron. Lett.
{\bf 11}, 133 (1985).
\bibitem{bruce} B. Allen and R. Caldwell, preprint (1994).
\bibitem{wald} R. Wald, {\em Quantum Field Theory in Curved
Spacetime} (University of Chicago Press 1994).
\bibitem{gorskiopen} K. M. Gorski, H. Ratra, N. Sugiyama and A. J.
Banday, preprint (1994).
\bibitem{rob}
R. R. Caldwell and A. Stebbins, preprint (1994).
\bibitem{turner} M. S.  Turner,  Phys Rev D {\bf 44}, 12 (1991).
\bibitem{misaolatest}  M. Sasaki, T. Tanaka and K. Yamamoto,
preprint  (1994).
\bibitem{lebedev} N. N. Lebedev,  C. R. (Doklady) Acad.  Sci.
URSS (N.S.)  {\bf 58}, 1007 (1946).
\bibitem{erdelyi} A. Erdelyi, Ed., {\sl Tables of Integral
Transforms}, (McGraw-Hill, 1953).
\bibitem{erdelyi2} A. Erdelyi, Ed.,
{\sl Higher Transcendental Functions, Volume II}
(McGraw Hill, 1953).

\end{thebibliography}
\end{document}